\documentclass{mn2e} 
\usepackage{epsf,times}
\usepackage{amsmath}
\usepackage{amssymb}
\usepackage{graphicx}
\usepackage{color}
\usepackage{natbib}
\usepackage{times}
\newcommand{\etal}{{et al}\/.}

\begin{document}

\title[FR II radio galaxies at low frequencies I]{FR II radio galaxies at low frequencies I: morphology, magnetic field strength and energetics}

\author[J.J.~Harwood \etal]{Jeremy J.\ Harwood$^{1}$\thanks{E-mail: jeremy.harwood@physics.org}, Judith H.\ Croston$^{2,3}$, Huib T. Intema$^{4,5}$, Adam J. Stewart$^{6}$, 
\newauthor Judith Ineson$^{2}$, Martin J.\ Hardcastle$^{7}$, Leith Godfrey$^{1}$, Philip Best$^{9}$, Marisa Brienza$^{1,8}$,
\newauthor Volker Heesen$^{2}$, Elizabeth K. Mahony$^{1}$, Raffaella Morganti$^{1,8}$, Matteo Murgia$^{11}$, 
\newauthor Emanuela Orr\'u$^{1,12}$, Huub R\"ottgering$^{4}$, Aleksandar Shulevski$^{1}$, Michael W. Wise$^{1,10}$
\vspace{2mm}
\\$^{1}$ASTRON, The Netherlands Institute for Radio Astronomy, Postbus 2, 7990 AA, Dwingeloo, The Netherlands
\\$^{2}$School of Physics and Astronomy, University of Southampton, Southampton SO17 1BJ, UK
\\$^{3}$Institute of Continuing Education, University of Cambridge, Madingley Hall, Madingley, CB23 8AQ, UK
\\$^{4}$Leiden Observatory, Leiden University, P.O. Box 9513, NL-2300 RA Leiden, The Netherlands
\\$^{5}$National Radio Astronomy Observatory, 1003 Lopezville Road, Socorro, NM 87801-0387, USA
\\$^{6}$Astrophysics, University of Oxford, Denys Wilkinson Building, Keble Road, Oxford OX1 3RH
\\$^{7}$School of Physics, Astronomy and Mathematics, University of Hertfordshire, College Lane, Hatfield, Hertfordshire AL10 9AB, UK
\\$^{8}$Kapteyn Astronomical Institute, University of Groningen, P.O. Box 800, 9700 AV, Groningen, The Netherlands
\\$^{9}$SUPA, Institute for Astronomy, Royal Observatory, Blackford Hill, Edinburgh, EH9 3HJ, UK
\\$^{10}$Astronomical Institute `Anton Pannekoek', University of Amsterdam, Postbus 94249, 1090 GE Amsterdam, The Netherlands
\\$^{11}$INAF, Osservatorio di Radioastronomia, Via della Scienza 5 - I-09047 Selargius (Cagliari)
\\$^{12}$Department of Astrophysics, Institute for Mathematics, Astrophysics and Particle Physics (IMAPP), Radboud University Nijmegen, P.O. Box 9010, 6500 GL\\Nijmegen, The Netherlands
}
\maketitle


\graphicspath{{./images/}}

\begin{abstract}

Due to their steep spectra, low-frequency observations of FR II radio galaxies potentially provide key insights in to the morphology, energetics and spectrum of these powerful radio sources. However, limitations imposed by the previous generation of radio interferometers at metre wavelengths has meant that this region of parameter space remains largely unexplored. In this paper, the first in a series examining FR IIs at low frequencies, we use LOFAR observations between 50 and 160 MHz, along with complementary archival radio and X-ray data, to explore the properties of two FR II sources, 3C452 and 3C223. We find that the morphology of 3C452 is that of a standard FR II rather than of a double-double radio galaxy as had previously been suggested, with no remnant emission being observed beyond the active lobes. We find that the low-frequency integrated spectra of both sources are much steeper than expected based on traditional assumptions and, using synchrotron/inverse-Compton model fitting, show that the total energy content of the lobes is greater than previous estimates by a factor of around 5 for 3C452 and 2 for 3C223. We go on to discuss possible causes of these steeper than expected spectra and provide revised estimates of the internal pressures and magnetic field strengths for the intrinsically steep case. We find that the ratio between the equipartition magnetic field strengths and those derived through synchrotron/inverse-Compton model fitting remains consistent with previous findings and show that the observed departure from equipartition may in some cases provide a solution to the spectral versus dynamical age disparity.

\end{abstract}


\quad\\

\section{Introduction}
\label{intro}

\begin{table*}
\caption{List of target sources, galaxy properties}
\label{targets}
\begin{tabular}{llccccc}
\hline
\hline
Name&IAU name&Redshift&5 GHz core flux&178 MHz flux&Spectral index&LAS\\
&&&density (mJy)&density (Jy)&(178 to 750 MHz)&(arcsec)\\
\hline
3C452&J2243$+$394&0.081&130&59.3&0.78&280\\
3C223&J0936$+$361&0.137&9.0&16.0&0.74&306\\
\hline
\end{tabular}

\vskip 5pt
\begin{minipage}{17.5cm}
`Name' and `IAU name' list the 3C and IAU names of the galaxies. `Spectral index' lists the low frequency spectral index between 178 to 750 MHz and `LAS' the largest angular size of the source. The `Redshift', `5 GHz core flux density', `178 MHz flux density', `Spectral index' and `LAS' column values are taken directly from the online version of the 3CRR database \citep{laing83} (http://3crr.extragalactic.info/cgi/database).
\end{minipage}
\end{table*}

\subsection{Radio galaxies}
\label{rgintro}

Radio galaxies can be broadly grouped into two categories: the centre-brightened \citet{fanaroff74} class I (FR I), and the edge brightened class II (FR II) galaxies. FR IIs generally consist of three key structures: jets, lobes and hotspots, and are the more powerful of the two classes with typical 1.4 GHz luminosities greater than $\sim$\,$10^{25}$ W Hz$^{-1}$  \citep{owen94}. The largest extent of jets and lobes in established FR IIs are commonly observed on scales of tens of kiloparsecs \citep{alexander87, konar06, machalski09} to megaparsecs \citep{mullin06, machalski08} in size and, as these lobes are in direct contact with their external environment, they are able to interact with both the intergalactic medium (IGM) and, if located in clusters, the intracluster medium (ICM). However, many unanswered questions remain about the underlying dynamics and energetics of these powerful radio sources. 

Radio galaxies are widely believed to play an important role in the evolution of galaxies and clusters (e.g. \citealp{croton06, bower06, fabian12, mcnamara12, morganti13, heckman14}). Their presumed ability to suppress star formation in models of galaxy evolution and to interact with their surrounding environment means that accurately determining the dynamics and energetics of these powerful outflows is vital if we are to understand the impact they have on how galaxies and clusters evolve over time. This work addresses important questions regarding the dynamics and energy content of two nearby, powerful FR II radio galaxies that are representative of the class. Specifically, we address the question of whether the lobes are overpressured relative to the external medium, and expanding supersonically. This question surrounding the dynamics also has a strong impact on the predicted scaling relation between jet power and radio luminosity \citep{godfrey15}, a key ingredient in studies of radio loudness and jet production \citep{sikora07} as well as quantifying radio mode feedback \citep{croton06}.

Limitations imposed by the previous generation of radio interferometers have meant that the brightness distributions of radio galaxies are currently poorly determined at metre wavelengths. Such observations are particularly important in the study of steep spectrum emission which is not detectable at higher frequencies, and in determining the energy content and distribution of the lowest energy electrons. With the LOw Frequency ARray (LOFAR; \citealp{haarlem13}) now fully operational, this is set to change. Consisting of $\sim$25000 high-band (HBA, 110 - 240 MHz) and low-band (LBA, 10 - 90 MHz) dipole antennas arranged into 46 stations spread throughout northern and central Europe, LOFAR is currently the world's largest connected interferometer, with baselines of up to $\sim$1500 km providing broad bandwidth, high resolution observations at low frequencies. LOFAR therefore now allows detailed low frequency spectral and morphological studies of radio galaxies on resolved spatial scales to be undertaken, and has already began to produce exciting new results both for well known active radio galaxies (e.g. 3C31, Heesen et al., in prep) and previously unknown remnant radio galaxies (e.g. \citealp{shulevski15, brienza16}). Such investigations are a key step if we are to accurately determine the impact of powerful radio galaxies on their environment.

\begin{table*}
\centering
\caption{Observation details}
\label{lofartargetobs}
\begin{tabular}{llcccccc}
\hline
\hline
Name&Array&Frequencies&Target&Flux&Calibrator&LOFAR project ID&Observation start date\\
&&(MHz)&TOS (mins)&calibrator&TOS (mins)&&\\
\hline
3C452&HBA Inner&110 -- 180&440&3C48&110&\verb|LC0_012|&27th August 2013\\
&LBA Outer&30 -- 80&589&3C48&589&\verb|LC0_012|&16th September 2013\\
3C223&HBA Inner&110 -- 180&429&3C196&78&\verb|LC0_012|&14th May 2013\\
&LBA Outer&30 -- 80&595&3C196&600&\verb|LC0_012|&23rd May 2013\\
\hline
\end{tabular}
\vskip 5pt
\begin{minipage}{15.1cm}
`Name' lists the 3C name of the galaxies and `Array' refers to the LOFAR array configurations used. `Frequencies' lists the frequency coverage of the observations. `Target TOS' lists the time on source at each frequency. `Calibrator TOS' lists the time on source at each frequency for the corresponding `Flux calibrator'. `LOFAR project ID' refers to the project identifier as used by the LOFAR archive search facility (http://lofar.target.rug.nl/).
\end{minipage}
\end{table*}

\subsection{The energetics and magnetic field of FR II radio galaxies}
\label{energeticsintro}

It has long been known that equipartition between the energy density of the relativistic particles in the lobes of radio galaxies and that of the magnetic field lies close to the minimum energy density required for the observed synchrotron-emitting plasma in radio galaxies \citep{burbidge56}. Assuming that these are the only two factors contributing to the energy density, the total energy content of a source can be determined solely via synchrotron emission at radio wavelengths; however, this has historically led to problems such as apparently underpressured radio lobes in seemingly expanding sources (e.g. \citealp{morganti88, hardcastle00, worrall00, harwood15}). A more robust calculation of the total energy content of FR II sources can be obtained where high resolution X-ray data are available. X-ray emission from the lobes of FR IIs is fairly common (e.g. \citealp{tashiro98, hardcastle00, isobe02, hardcastle04b}) and is thought to be a result of the inverse-Compton process upscattering cosmic microwave background photons (iC/CMB) (e.g. \citealp{kataoka05, hardcastle05}). As the iC/CMB emissivity does not depend on the magnetic field of the source, modelling of the combined synchrotron/inverse-Compton spectrum (constrained by measurements at radio and X-ray energies respectively) can provide a more robust measure of the magnetic field strength (e.g. Croston et al., 2004, 2005).

One key assumption required for these models is the spectrum at low frequencies, as the magnetic field calculation involves scaling from the synchrotron emitting electrons with $\gamma \sim 10^{4}$ to the inverse-Compton electrons with $\gamma \sim 10^{3}$. Whilst integrated flux densities at these wavelengths have long been available, the limited number of data points has meant that the spectrum of this emission must be assumed. As the spectrum of these sources is always relatively steep ($\alpha > 0.5$)\footnote{We define the spectral index such that $S \propto  \nu^{-\alpha}$} any significant change in this assumption can greatly impact on the derived energy content of the source. LOFAR observations therefore provide the ability to constrain these models and evaluate the previously used assumptions about low-frequency spectrum and magnetic field strength.

\subsection{Outstanding questions addressed in this paper}
\label{questions}

In this paper, the first in a series examining FR II radio galaxies at low frequencies, we use LOFAR observations to explore the morphology, magnetic field strength and energetics of two powerful radio sources. We address three primary questions:

\begin{enumerate}
\item How does the morphology of FR II radio galaxies at LOFAR frequencies compare to previous studies at higher frequencies?\\
\item What is the magnetic field strength of FR II radio galaxies and does it agree with that derived from equipartition?\\
\item Do improved constraints placed on the low-energy electron distribution change our understanding of the energetics of FR II radio galaxies?
\end{enumerate}

In Section \ref{method} we give details of target selection, data reduction and the analysis undertaken. Section \ref{results} presents our results and in Section \ref{discussion} we discuss these findings in the context of the aims outlined above. Throughout this paper, a concordance model in which $H_0=71$ km s$^{-1}$ Mpc$^{-1}$, $\Omega _m =0.27$ and $\Omega _\Lambda =0.73$ is used \citep{spergel03}.

\section{Data Reduction and Spectral Analysis}
\label{method}

\subsection{Target selection and observations}

Thirteen targets were observed as part of the surveys Key Science Project (KSP), cycle-0 nearby AGN proposal covering various stages in the life-cycle of radio-loud active galaxies. Of this sample, two nearby powerful radio galaxies for which complementary X-ray observations were available, 3C223 and 3C452, were suitable for the study of active FR II sources (Table \ref{targets}). To ensure good UV coverage and sensitivity, 10 hour observations were made of each target at HBA and LBA frequencies using the core and remote Dutch stations. At the time of these observations, this provided baselines out to 100 km which, at the lowest HBA frequency (110 MHz), give a resolution of $\sim$7 arcseconds.

Suitable flux calibrators were selected for each target from a list of reference sources \citep{scaife12} which, for HBA observations, were interlaced with the target observations at 11 minute intervals. At LBA frequencies, where correcting for changes in the atmosphere and side lobe effects on short time scales is extremely important, the calibrator source was observed concurrently with the target by using some of the available bandwidth to form a second beam on the sky which constantly monitors the calibrator source. A summary of the observational setup is given in Table \ref{lofartargetobs}.

\subsection{Data reduction}
\label{datareduction}

The calibration of the data used in this paper follows the guidelines set out in the standard imaging cookbook\footnote{https://www.astron.nl/radio-observatory/lofar/lofar-imaging-cookbook/}; however, as LOFAR data reduction is a relatively new process, we summarize here some of the key procedures and LOFAR specific functions that have been written for the handling of these low-frequency data.

One key difference in the dipole setup of LOFAR compared to observations using parabolic dishes is the need for the removal of bright radio sources. As LOFAR always sees the entire sky, the 5 brightest radio sources (Cassiopeia A, Cygnus A, Hydra A, Taurus A and Virgo A) must be subtracted from the data if they are expected to interfere heavily with a given observation. These so-called A-team sources, the properties of which have been well characterized, are therefore removed in a process known as demixing prior to the data being made available to the end user. However, if this demixing is performed when interference from A-team sources is not present, the data quality can deteriorate. Prior to observation, simulations were therefore run in order to determine if the demixing of A-team sources was required. We found that at both LBA and HBA frequencies, Cygnus A and Cassiopeia A heavily affect the observations, and demixing of these sources was undertaken. To reduce the datasets to a manageable size, each sub-band was also averaged down to four, 48.8 kHz channels and to 5 (HBA) and 10 (LBA) second integration times by the LOFAR support staff prior to the data being made available.

\begin{figure*}
\centering
\includegraphics[angle=0,height=7cm]{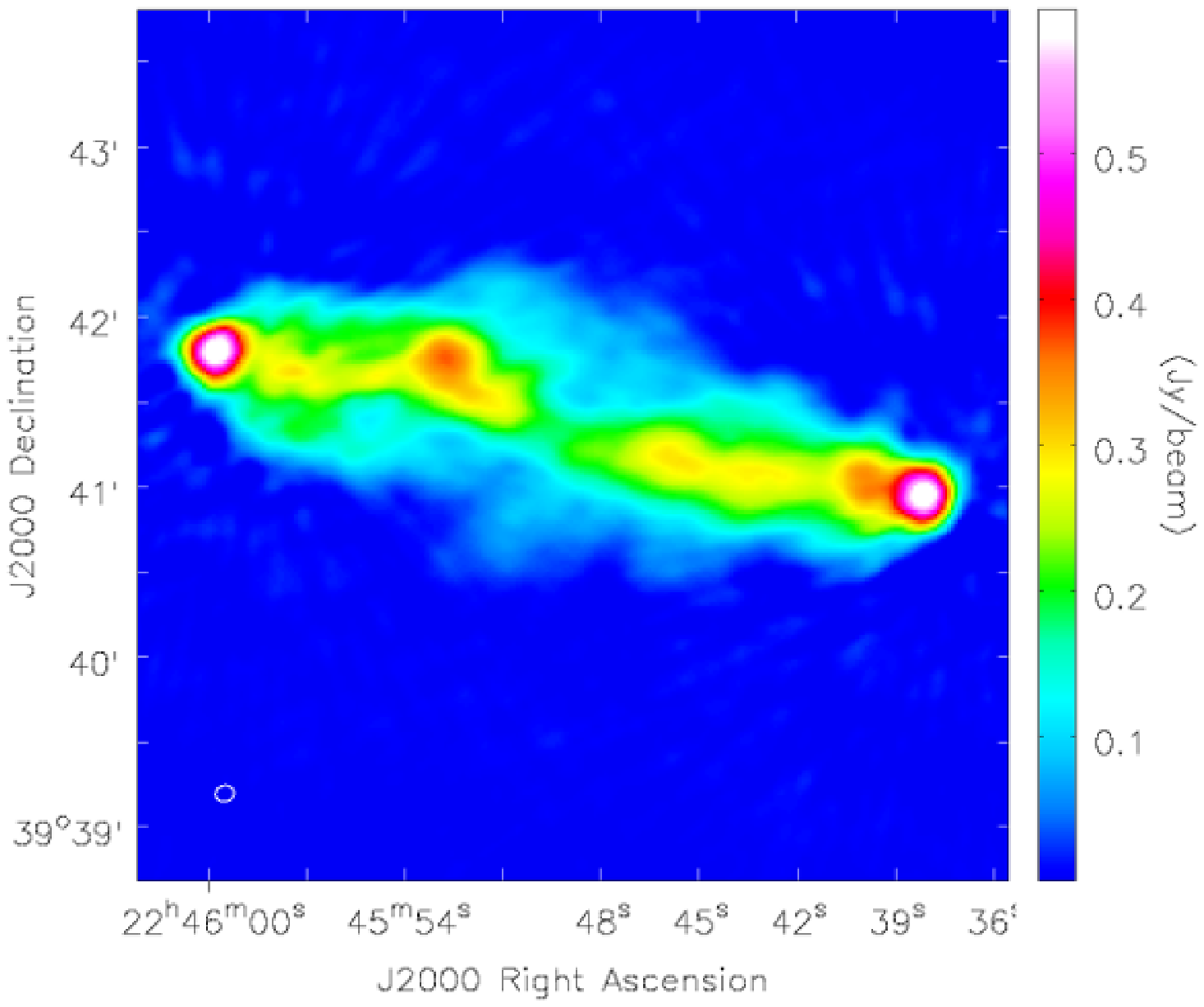}
\includegraphics[angle=0,height=7cm]{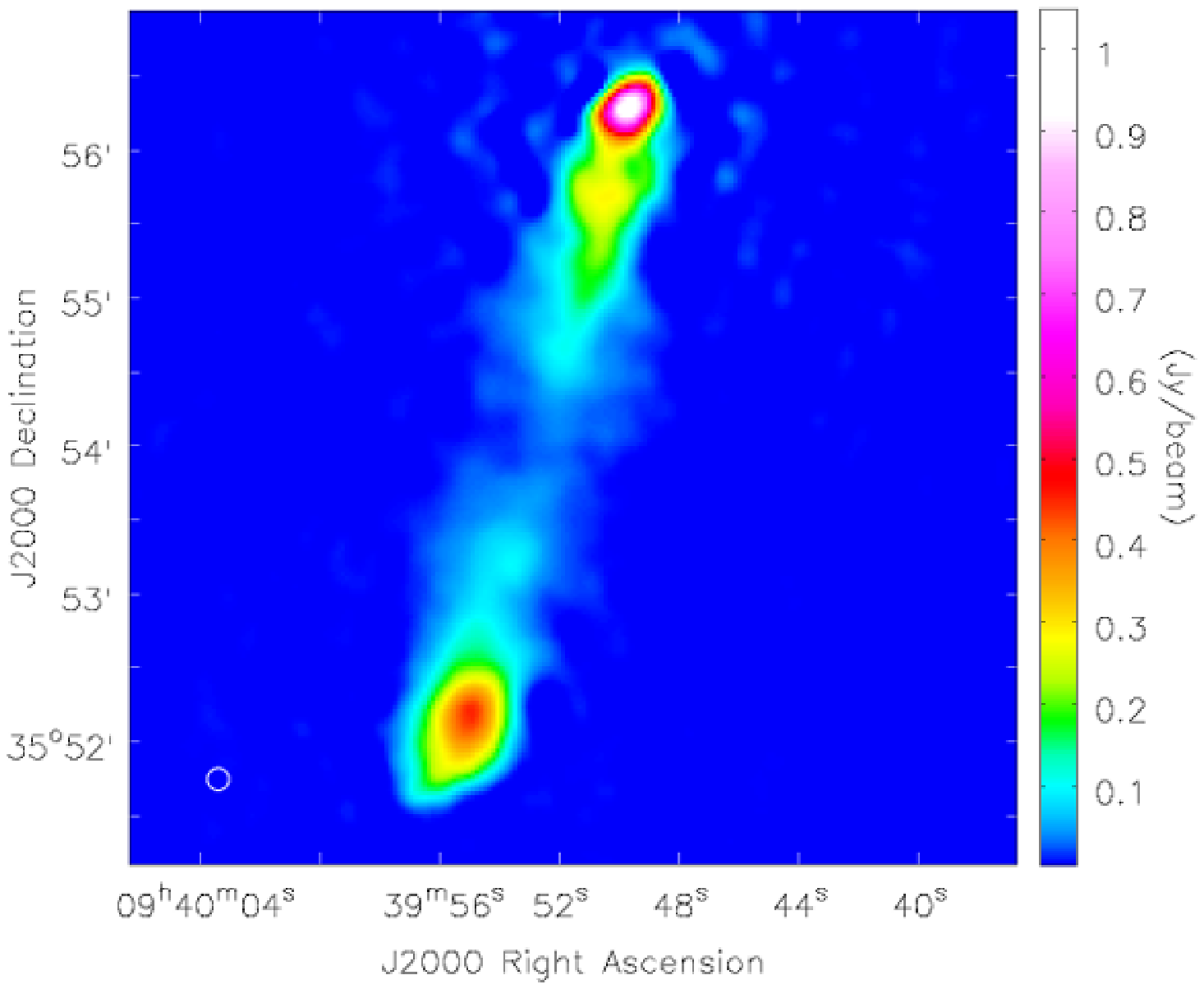}\\
\caption{Radio maps of 3C452 at 138 MHz (left) and 3C223 at 147 MHz (right), imaged using multiscale CLEAN in \textsc{casa}. The off-source RMS noise is $0.45$ mJy beam$^{-1}$ for 3C452 and $0.48$ mJy beam$^{-1}$ for 3C223. The restoring beam is indicated in the bottom left corner of the image.}
\label{lofarexamplemaps}
\end{figure*}

Once complete, the 3C223 and 3C452 data were downloaded from the archive to the LOFAR cluster at Southampton University and the ASTRON CEP3 cluster respectively. Due to the high volume of data produced by LOFAR (even after averaging), a complete manual reduction of observations is not desirable, and so initial calibration was carried out using the observatory pipeline \citep{heald10}. Within this pipeline, flagging of visibilities affected by radio frequency interference (RFI) was first performed on the demixed and averaged data using the AOFlagger of \citet{offringa10}. Amplitude calibration and phase solutions were then determined from the calibrator sources (Table \ref{lofartargetobs}) using BlackBoard Selfcal (\textsc{bbs}, \citealp{pandey09}), which accounts for variations of the LOFAR station beams, and the solutions applied. Models of the fields of view were then created using the Global Sky Model (GSM; \citealp{scheers11}) function which uses the VLA Low-frequency Sky Survey (VLSS), Westerbork Northern Sky Survey (WENSS) and NRAO VLA Sky Survey (NVSS) surveys \citep{condon94, condon98, rengelink97, cohen07} to determine the source properties. A phase only calibration was then carried out using \textsc{bbs}, all solutions were transferred to the target source, the array beam applied, and a further round of automated flagging performed on the corrected data. Finally, the individual subbands of the LOFAR observations were concatenated using the New Default Pre-Processing Pipeline (\textsc{ndppp}) package and a manual inspection of the data made. To obtain a reasonable number of data points across the frequency space, but also maintain a good signal to noise ratio, the data were concatenated into 4-MHz bands for 3C452 and 7-MHz bands for 3C223 at HBA frequencies. At LBA frequencies, both observations were severely affected by the ionosphere and so were unsuitable for our analysis, with the exception of a single 12-MHz band for 3C223 centred at 52 MHz. Any remaining RFI was then flagged manually using \textsc{casa} in the standard manner. 

In order to further improve the image quality, direction independent phase only self-calibration was then performed on each dataset. \textsc{casa} was chosen as the preferred imaging and self-calibration tool in order to utilise the multi-frequency synthesis (MFS) algorithm with two Taylor series terms (nterms = 2; \citealp{rau11}), which scales the flux by a spectral index value fitted over the observed bandwidth, and multi-scale cleaning which were not yet available to the LOFAR specific AWImager \citep{tasse13}. This has the advantage of both enhancing the image fidelity and significantly reducing the computing time required. While this removed the ability to correct for the attenuation of the LOFAR primary beam, from figure 21 of \citet{haarlem13} we see that at 150 MHz the sensitivity at a field radius of 30 arcminutes remains $>$95 per cent of the peak response. The fact that the sources are located at the centre of the field of view and have a largest angular size of only around 5 arcminutes means that any losses due to the sensitivity drop off could safely be ignored. Unlike in the high frequency regime, the LOFAR self calibration strategy involves including only the core LOFAR stations and iteratively moving outwards to include longer baseline. This is due to the remote stations (i.e. the longest baselines) not being on the same clock leading to potential time offsets, and ensures that phase coherence is maintained. Due primarily to RFI, instrumental and ionospheric effects (particularly at the edge of the observing frequency), at HBA wavelengths only 6 of the 7-MHz bands for 3C223 and 10 of the 4-MHz bands for 3C452 between $\sim$116 and 160 MHz were suitable for final imaging.

\begin{figure}
\hspace{3.5mm}
\includegraphics[angle=0,width=7.5cm]{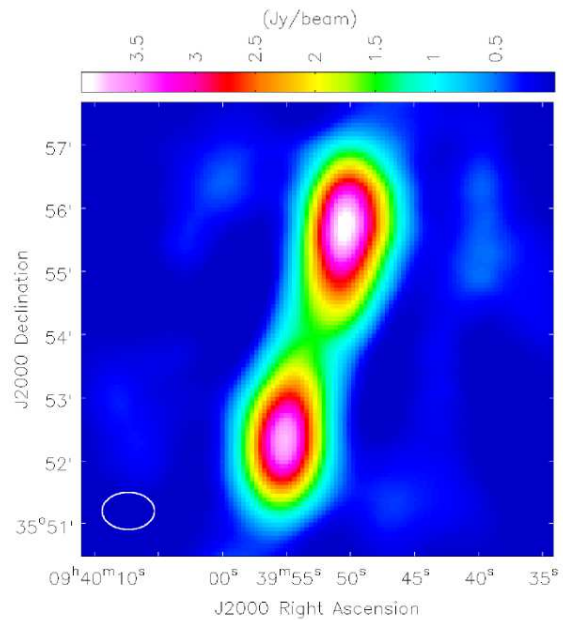}
\caption{LOFAR LBA radio map of 3C223 at 51.6 MHz with an off-source RMS noise of $12.4$ mJy beam$^{-1}$. The restoring beam is indicated in the bottom left corner of the image.}
\label{3c223lbamap}
\end{figure}

Once self calibrated, the images were mapped at the resolution appropriate to the lowest LBA and HBA frequencies with a cell size one fifth of that value. Although within this paper we are only interested in the integrated fluxes of the sources, these parameters ensure that the images are well matched in resolution to allow a full spectral ageing analysis to be performed in the second paper in this investigation. A summary of parameters used for the imaging described in this section is shown in Tables \ref{lofarimagrprmsHBA} and \ref{lofarimagrprmsLBA}. HBA images were also created using the full bandwidth centred at a frequency of 147 and 138 MHz for 3C223 and 3C452 respectively. The resulting HBA and LBA images are shown in Figures \ref{lofarexamplemaps} and \ref{3c223lbamap}.

\subsection{Synchrotron/inverse-Compton model fitting}
\label{modelfitting}

As was discussed in Section \ref{energeticsintro}, it is a common assumption that the relativistic particles in the lobes of radio galaxies and the magnetic field are the only contributing factors in determining the energy density of the lobes; however, there is no \emph{a priori} reason to believe that the total energy content (the sum of the energy density in the radiating particles, the magnetic field, and any non-radiating particles such as protons) is not significantly higher than the minimum energy condition. This has made estimates of the energy budget based on radio observations alone highly uncertain.

The availability of X-ray data for our target sources reduces this problem by allowing a determination of the magnetic field strength to be made free from the assumption of equipartition and minimum energy. The total losses suffered by particles emitting at radio wavelengths are a combination of inverse-Compton and synchrotron losses, which are proportional to the magnetic field strength in the lobes and the equivalent field of the CMB such that $B = \sqrt{B_{lobe}^{2} + B_{CMB}^{2}}$. The equivalent field of the CMB scales with redshift as $B_{CMB} = 0.318 (1 + z)^{2}$, hence the energy losses of the particles due to inverse-Compton scattering at low redshifts only becomes important at low frequencies.

Conversely, the energy gained by the CMB photons is given by \citep{longair11} \begin{equation}\label{cmbgains}\frac{dE}{dt} = \frac{4}{3} \left(\frac{E}{m_{e} c^{2}} \right)^{2} \sigma_{T} u_{CMB} \end{equation} where $u_{CMB}$ is the energy density of the CMB. The X-ray inverse-Compton emission is therefore dependent on only the supply of photons that will be up-scattered, in this case the CMB, and the energy of the scattering particles. Provided that the spectrum of a source is well known at radio wavelengths (particularly at low frequencies), the 1 keV inverse-Compton X-ray emission can be combined with the magnetic field dependent radio synchrotron emission to give constraints on the magnetic field strength of the source.

For FR IIs where the contribution from non-radiating particles is thought to be negligible \citep{wardle98, homan99, croston05, konar13}, and with the magnetic field now known, a more robust estimate of the energy density of the synchrotron-emitting plasma, hence total energy content of the lobes, can also be made. To investigate the energetics of our two sources, we therefore refitted the synchrotron/inverse-Compton models used by Croston et al. \citeyear{croston04} and \citeyear{croston05}, constrained by the new radio observations presented here. In addition to the LOFAR measurements, flux densities at 330 MHz and 1.4 and 8 GHz were included in the synchrotron/inverse-Compton modelling which, along with the X-ray flux, were taken to be the values measured by Croston et al. \citeyear{croston04} and \citeyear{croston05}. Archival low-frequency measurements were also included to extend our frequency coverage down to around 10 MHz for 3C452 and 20 MHz for 3C223. In order to constrain the spectrum of 3C452 at LBA frequencies we also include the integrated flux value presented by \cite{kassim07} who use the Jansky Very Large Array (JVLA) at 73.8 MHz assuming a 5 per cent uncertainty in the flux calibration. A summary of the archival data used is given in Table \ref{archivalmapdetails}.

In order to ensure the results are directly comparable to those of \citet{croston05}, 3C452 uses the total flux for the model fitting as its morphology means there is no clear division between the two lobes. For 3C223, the flux of each lobe was determined for the unresolved archival data using the ratio between the northern and southern lobes of 49:51 derived by \citet{croston04} who used the average of the fraction of the total flux in each lobe in the resolved images at 1.5 and 8 GHz.

The \textsc{synch} code of \citet{hardcastle98} was used to fit a two component model consisting of a synchrotron spectrum constrained by low-frequency radio measurements, and an inverse-Compton spectrum constrained by X-ray observations. We applied the model presented by \citet{croston05} which used an electron distribution described by a single power law and bounded by the high and low energy cutoffs given in Table \ref{lofarsynchparams}. The radio lobes were modelled as cylinders and we assumed a minimum energy cutoff where $\gamma_{min} = 10$. This value is constant with that used by \citet{croston04, croston05} and was based on observations of hotspots in FR IIs where $\gamma_{min} \sim$ 100 to 1000 (e.g. \citealp{carilli91, hardcastle98}) which, after adiabatic expansion (e.g. \citealp{godfrey09}), reduces to a $\gamma_{min}$ of approximately 10 for the lobes.

In theory, particles are shock accelerated in the hotspot regions of FR IIs to a power law energy distribution such that \begin{equation}\label{initialpowerlaw}N(E) = N_0 E^{-\delta}\end{equation} The model parameter which describes the observed spectrum produced by this power law distribution of shock accelerated particles (the injection index) is directly related to the $\delta$ term by \begin{equation}\label{alphainject}\alpha_{inj} = \frac{\delta - 1}{2}\end{equation} For steep spectrum sources ($\delta > 2$) where $\gamma_{max} >> \gamma_{min}$, a change in $\alpha_{inj}$ will have a significant impact on the energy stored at low frequencies scaling as \begin{equation}\label{injindexdep}U_{e} \propto \frac{ \gamma_{min}^{2-\delta} }{\delta - 2}\end{equation} where $U_{e}$ is the total energy contained in the relativistic electrons. For the special case where $\delta = 2$ ($\alpha_{inj}=0.5$), the dependence on $\gamma_{max} / \gamma_{min}$  becomes logarithmic and so the impact on the total energy content is reduced.

The injection index was determined empirically from the best fitting low-frequency spectral index of the LOFAR images and the measurements of Croston et al. \citeyear{croston04} and \citeyear{croston05}. As losses due to spectral ageing should be low at these wavelengths (discussed further in Section \ref{energeticsdicussion}), this should provide a robust estimate of the initial power law distribution.

\section{Results}
\label{results}

\subsection{Data quality and integrated flux}
\label{integratedflux}

Before an analysis of these sources can be undertaken (particularly in the case of new instruments such as LOFAR) we must first consider the reliability of the calibration and any errors associated with the data. Comparison to previous studies provides a good check of our flux calibration. Using the integrated flux values and spectral index for 3C223 presented by \citet{orru10} who use VLA observations at 74 and 327 MHz and extrapolating to LOFAR LBA and HBA frequencies, we find total integrated flux values of 37.4 Jy at 51.6 MHz and 17.9 Jy at 161 MHz. As a secondary check, the same process can be carried out using the integrated fluxes tabulated by \citet{laing80} at 84 MHz and 178 MHz. These give similar values of 36.9 Jy at 51.6 MHz and 17.0 Jy at 161 MHz. Following the same process for 3C452 using the values of \citet{nandi10} at 153 MHz and the spectral index given by \citet{hardcastle04b}, we find that at 152 MHz the integrated flux is 81.6 Jy. Comparing these values to Table \ref{lofarmapdetails}, and assuming a calibration error of 10 per cent at HBA and 15 per cent LBA frequencies due to uncertainties in the flux scale and global beam model \citep{scaife12, weeren14}, we find our measurements are in good agreement with the literature. We note that for both sources the HBA in-band spectral index is flatter than one might expect given the overall spectrum between 50 MHz and 1.4 GHz. Uncertainties associated with the LOFAR HBA beam, particularly for measurements taken away from the central frequencies, mean that the in-band spectrum is currently still unreliable; however, the overall spectral index remains well within the stated calibration errors.

While the thermal noise measured in the images for both sources at HBA frequencies is relatively low ($\lesssim$ 1 mJy beam$^{-1}$, Table \ref{lofarmapdetails}), the noise close to the sources is significantly higher by a factor of around 5$-$10 times than when measured in a blank region of the sky. This increased noise is observed in both sources close to the hotspots and is therefore likely to be an issue related to dynamic range limitations in these regions. This is particularly prominent in the northern lobe of 3C223 where the dynamic range is greatest. For the purposes of our analysis, the increased RMS close to the source remains fractionally low compared to the integrated flux ($\lesssim$1 per cent at HBA frequencies) and should therefore not impact significantly upon our results.

\begin{table}
\centering
\caption{Summary of HBA imaging parameters}
\label{lofarimagrprmsHBA}
\begin{tabular}{llcl}
\hline
\hline
Parameter&\textsc{casa} Name&Value&Units\\
\hline
Polarization&\textsc{stokes}&I&\\
Image Size&\textsc{imsize}&4096 4096&pixels\\
Cell Size&\textsc{cell}&1.4 1.4&arcsec\\
Weighting&\textsc{robust}&-0.5&\\
Beam size&\textsc{restoringbeam}&7.0 7.0&arcsec\\
Multiscale&\textsc{multiscale}&[0, 5, 15, 45]&pixels\\
\hline
\end{tabular}
\vskip 5pt
\begin{minipage}{8cm}
`Parameter' refers to the imaging parameter used in making of radio maps within this chapter. `{\sc casa} Name' refers to the {\sc casa} parameter taking the value stated in the `Values' column.
\end{minipage}
\end{table}

\subsection{Morphology}
\label{morphology}

From Figure \ref{lofarexamplemaps} we see that both 3C223 and 3C452 have a classical FR II morphology consisting of edge brightened, diffuse lobe emission with clear hotspots towards the end of the source. The morphology of 3C223 closely matches that of previous studies (e.g. \citealp{leahy91a, orru10}) with previously known structure at comparable resolutions being successfully recovered. While for 3C223 the observed morphology is what one would expect for a classical FR II, 3C452 proves to be a more interesting case.

It was recently suggested by \citet{sirothia13} that 3C452 may be a so-called double-double (DDRG; \citealp{schoenmakers00}). DDRGs contain two pairs of radio lobes resulting from two separate episodes of AGN activity: a pair of faint outer lobes from a previous active phase, and a pair of bright, inner lobes from the current outburst of activity. Panel A of Figure \ref{ddrgcompimage} shows the image presented by \citet{sirothia13}, where two pairs of these lobe-like structures are clearly visible, and the LOFAR image presented within this paper with a comparable field of view. While the active inner lobes of 3C452 recover the same structure in both images, it is immediately clear that the outer remnant lobes are not observed in the LOFAR observations. We discuss the possible causes of this missing emission in Section \ref{noddrg}.

\begin{table}
\centering
\caption{Summary of LBA imaging parameters}
\label{lofarimagrprmsLBA}
\begin{tabular}{llcl}
\hline
\hline
Parameter&\textsc{casa} Name&Value&Units\\
\hline
Polarization&\textsc{stokes}&I&\\
Image Size&\textsc{imsize}&4096 4096&pixels\\
Cell Size&\textsc{cell}&7.0 7.0&arcsec\\
Weighting&\textsc{robust}&-0.5&\\
Beam size&\textsc{restoringbeam}&49.7 35.4&arcsec\\
Position Angle&\textsc{restoringbeam}&89.8&degrees\\
Multiscale&\textsc{multiscale}&[0, 5, 15, 45]&pixels\\
\hline
\end{tabular}
\vskip 5pt
\begin{minipage}{8.0cm}
`Parameter' refers to the imaging parameter used in making of radio maps within this chapter. `{\sc casa} Name' refers to the {\sc casa} parameter taking the value stated in the `Values' column.
\end{minipage}
\end{table}

\subsection{Energy density and magnetic field strength}
\label{energertics}

Within errors, the integrated LOFAR fluxes of both sources agree well with the original observations at other frequencies made by \citet{croston04, croston05}; however, the now well constrained spectrum of the emission at very low frequencies provides some potentially interesting differences with respect to the total energy content of the lobes of these sources. The integrated spectrum at these frequencies is much steeper than previously expected with an index of 0.85 for 3C452 and 0.71 for 3C223, implying a greater amount of energy is contained in the low-energy electron population than previously thought.

From Figure \ref{synchicfit}, which shows the results of the synchrotron/inverse-Compton model fitting under the assumption that this steep spectral index is representative of the underlying initial electron energy distribution (Table \ref{lofarsynchparams}), one finds that in order for the model to match the observed 1 keV inverse-Compton flux, an energy density of $1.2 \times 10^{-12}$  J m$^{-3}$ is required for 3C452 and of $2.8 \times 10^{-13}$ and $3.2 \times 10^{-13}$ J m$^{-3}$ are required for the northern and southern lobes of 3C223 respectively (Table \ref{lofaricres}). Comparing these values to those presented by \citet{croston04, croston05} we find a change in the total lobe energy density by a factor of 5.0 for 3C452 and by 2.3 in the northern and 2.0 in the southern lobes of 3C223, a significant increase over previous estimates.

One consideration that should be made when determining the total energy content of the lobes is the uncertainty associated with $\gamma_{min}$. While this value remains unconstrained, \citet{croston05} show that even if $\gamma_{min} = 1000$ in the lobes, the effect on the total energy content is small, varying by only a factor of $\approx 2$ for the injection index values used here. This is due to the electron energy spectrum's normalization increasing to maintain equipartition, so offsetting the reduction at the lowest energies and remaining consistent within the measurement errors. Therefore while a significant shift in $\gamma_{min}$ could cause the total energy content of 3C223 to be in closer agreement with previous estimates due to the injection index of $\alpha_{inj} = 0.5$ assumed by \citet{croston04} scaling logarithmically (Section \ref{modelfitting}) rather than by Equation \ref{injindexdep}, this is unlikely to be the case for 3C452 where the difference is much larger.

The magnetic field strength is also affected by these revised estimates of the initial electron energy distribution. Comparing the results of Table \ref{lofaricres} to those of \citet{croston04, croston05} we see an increase in the magnetic field strength of $\sim$60 per cent for both 3C452 and 3C223. However, as a natural consequence the increased electron energy content at low frequencies also impacts on the equipartition value, causing it to increase from $\sim$0.50 to 0.87 nT in 3C452, and from $\sim$0.35 to 0.45 nT in 3C223. The values derived from X-ray constraints therefore remain lower than the equipartition value as originally shown by \citet{croston04, croston05}. We discuss the implications of the steeper than expected low frequency spectrum and the potential implications of the derived values further in Section \ref{energeticsdicussion}.

\begin{table}
\centering
\caption{Summary of archival data by frequency}
\label{archivalmapdetails}
\begin{tabular}{lcccc}
\hline
\hline
Source&Frequency&Integrated flux&Uncertainty&Reference\\
&(MHz)&(Jy)&(Jy)&\\
\hline
3C452&10&396.0&108.0&1\\
&22.3&331.2&21.7&1\\
&26.3&288.0&18.0&1\\
&38&236.0&70.8&1\\
&73.8&142.0&8.6&2\\
&86&114.0&3.0&1\\
&178&59.30&2.97&3\\
&1400&10.54&0.27&3\\
&8350&1.86&0.04&3\\

3C223&26.3&83.0&7.0&1\\
&38&47.95&4.30&1\\
&86&27.6&1.1&1\\
&178&16.0&0.8&1\\
&330&11.7&0.4&4\\
&1477&3.49&0.07&5\\
&8350&0.89&0.05&6\\

\hline

\end{tabular}

\vskip 5pt
\begin{minipage}{8.5cm}
`Frequency' refers to the frequency at which measurements were made for each `Source'.  The corresponding fluxes are listed in the `Integrated flux' column with errors in the measurements shown in the `Uncertainty' column. The `Reference' column gives the origin of each value, denoted as follows: (1) \citet{laing80}; (2) \citet{kassim07}; (3) \citet{croston05}; (4) \citet{croston04}; (5) \citet{leahy91a}; (6) \citet{hardcastle98a}.
\end{minipage}
\end{table}

\section{Discussion}
\label{discussion}

The morphology and energetics of powerful radio galaxies play an important role in our understanding of the life-cycle of radio galaxies and the impact they have on their environment. Properties such as the total lobe energy content and magnetic field strength are key parameters which must be reliably determined if we are to derive properties such as source ages, lobe pressures and, ultimately, the total energy output of radio galaxies which can impact on galaxy evolution as a whole. 

In this paper we have presented results which explore the low-frequency morphology, energetics and integrated spectrum of two powerful radio galaxies, 3C452 and 3C223, which suggest a disparity from previous investigations. In the following section we examine possible causes for our findings and their implications for our understanding of the FR II population.

\begin{table}
\centering
\caption{Summary of synchrotron/inverse-Compton model parameters}
\label{lofarsynchparams}
\begin{tabular}{llcll}
\hline
\hline
Source&Parameter&Value&Units&Comments\\
\hline
3C452&Length, Radius&289.2, 89.0&arcsec&Total source\\
&$\delta$&2.70& &$\delta = 2\alpha_{inj} + 1$\\
&$E_{min}$&$5 \times 10^{6}$&eV&$\gamma \approx 10$\\
&$E_{max}$&$1 \times 10^{11}$&eV&$\gamma \approx 1\times10^{6}$\\
3C223&Length, Radius&157.5, 21.5&arcsec&Northern lobe\\
&Length, Radius&152.0, 25.8&arcsec&Southern lobe\\
&$\delta$&2.42& &$\delta = 2\alpha_{inj} + 1$\\
&$E_{min}$&$5 \times 10^{6}$&eV&$\gamma \approx 10$\\
&$E_{max}$&$1 \times 10^{11}$&eV&$\gamma \approx 1\times10^{6}$\\
\hline
\end{tabular}
\vskip 5pt
\begin{minipage}{8.5cm}
`Parameter' refers to the parameter name, where $E_{min}$ and $E_{max}$ are the assumed minimum and maximum energy of the electron distribution respectively and are taken directly from \citet{croston04, croston05}. `Length, Radius' are the source dimensions assuming a cylindrical geometry and $\delta$ is the electron energy power law index given by $\delta = 2\alpha_{inj} + 1$ where $\alpha_{inj}$ is the injection index of the source. Note that for 3C452 we use the total size of the source, as opposed to the individual lobes, to allow direct comparison to the results of \citet{croston04, croston05}.
\end{minipage}
\end{table}

\begin{figure*}
\centering
\hspace{5.5mm}
\includegraphics[angle=0,width=7.5 cm]{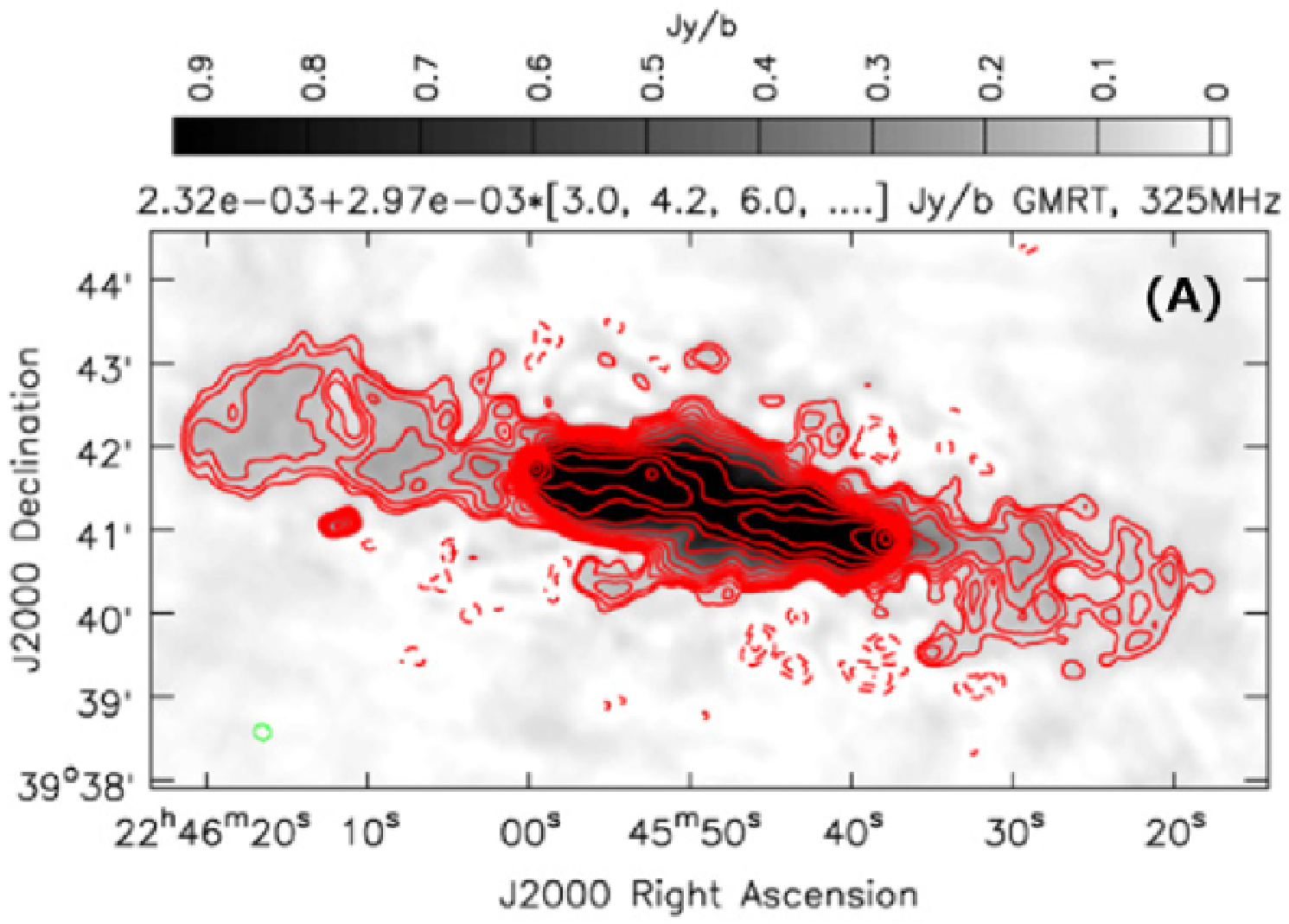}
\includegraphics[angle=0,width=9.5cm]{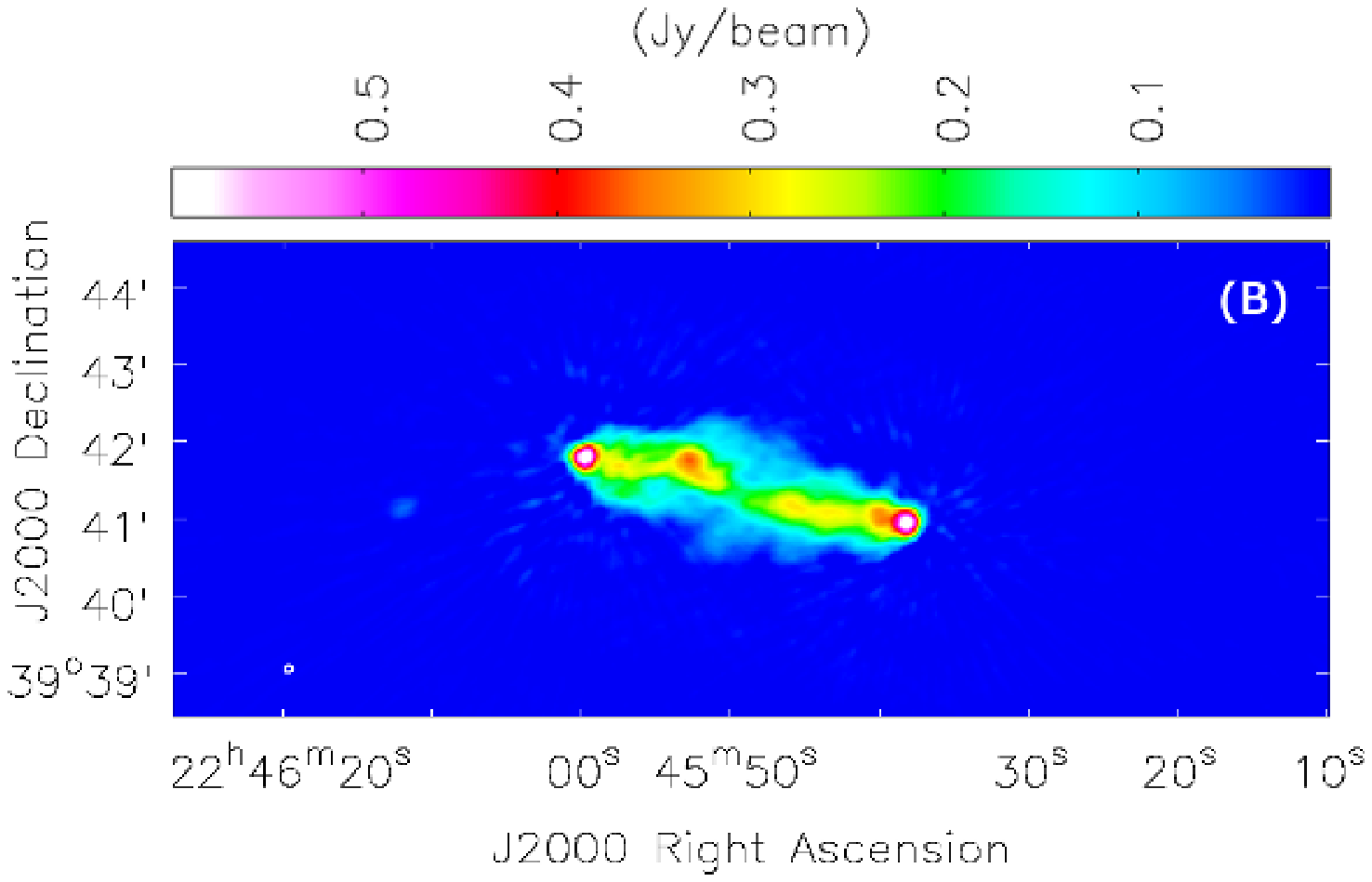}\\
\includegraphics[angle=0,width=8.8cm]{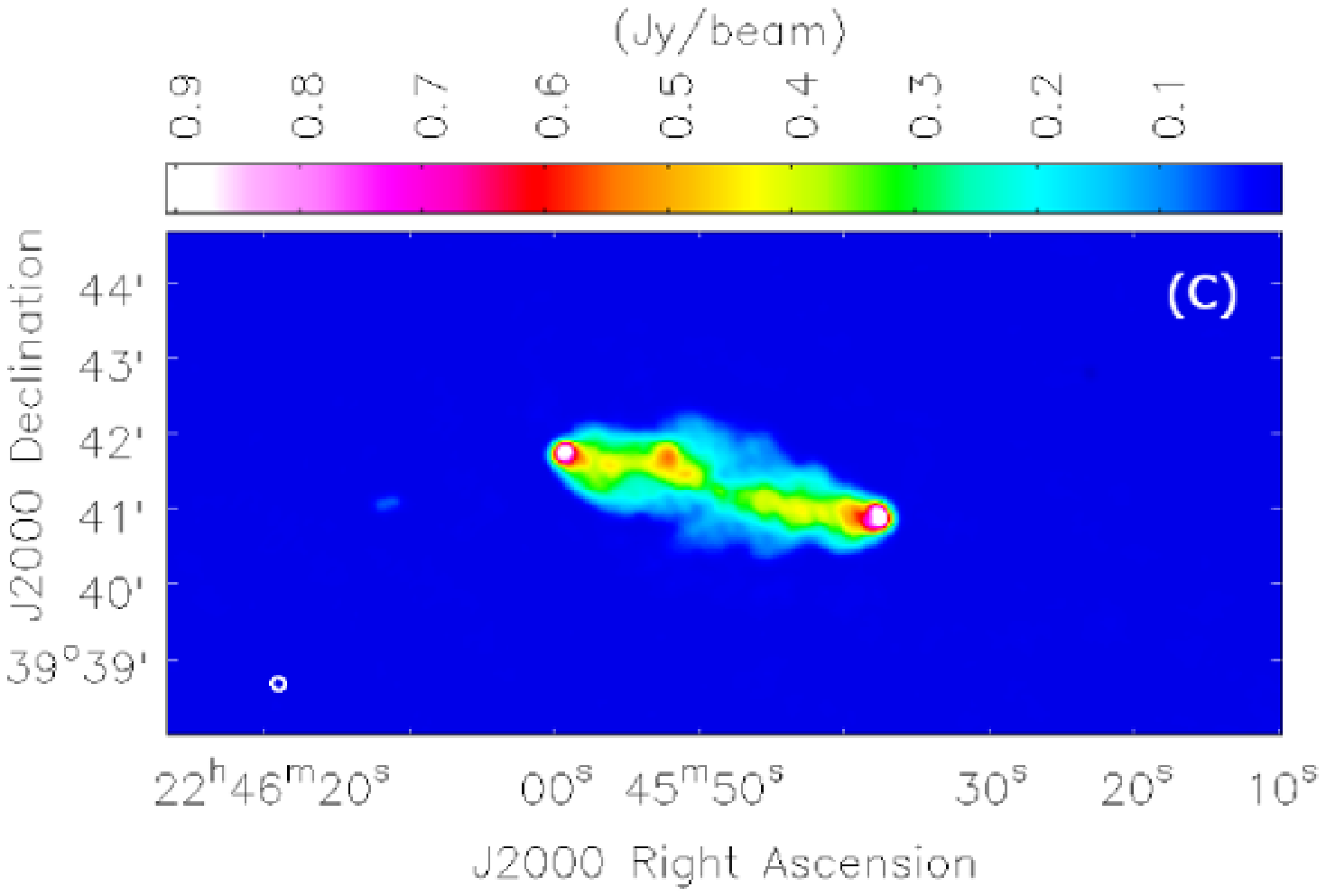}
\includegraphics[angle=0,width=8.8cm]{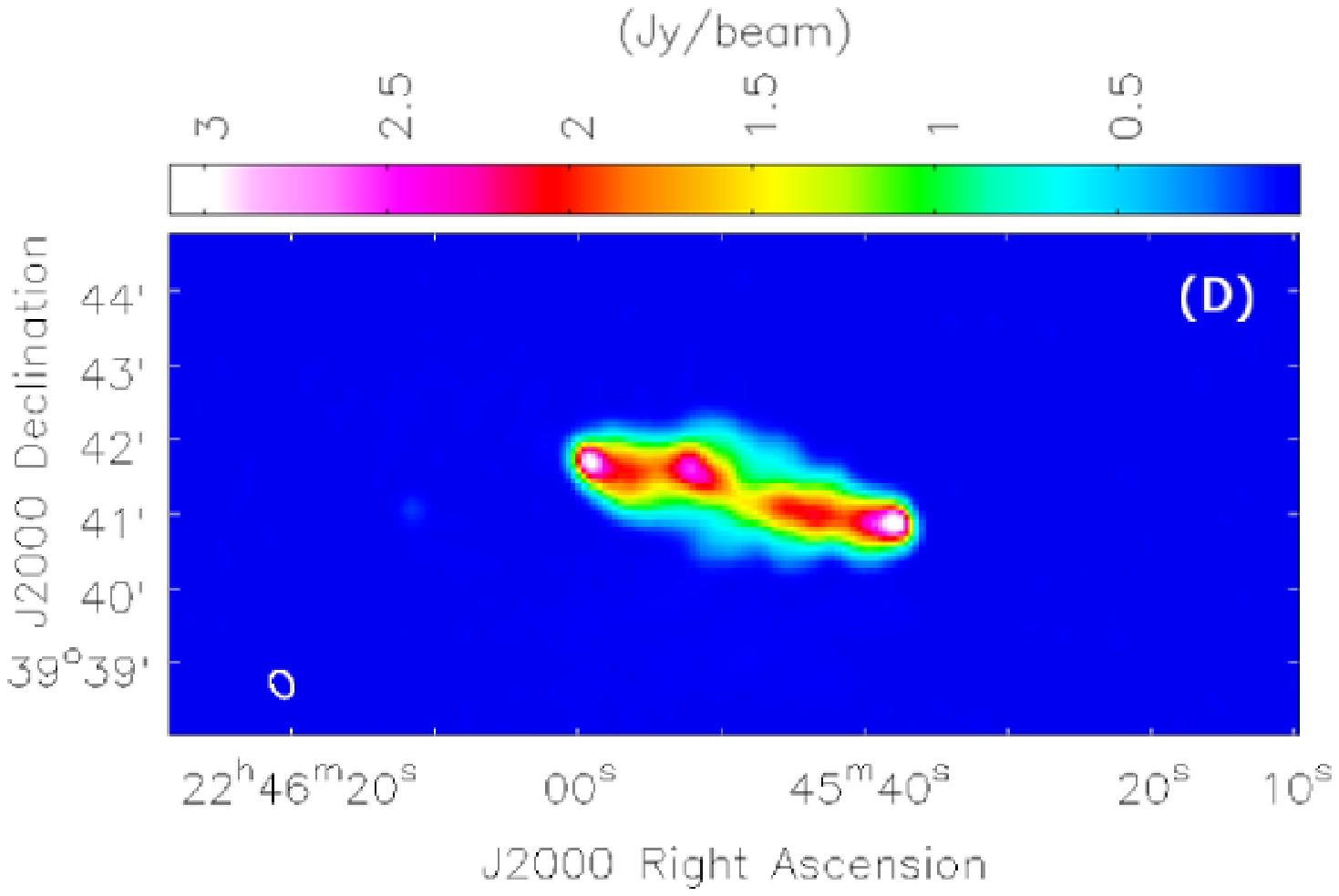}\\
\caption{Radio maps of 3C452 and the surrounding environment. (A) GMRT image at 325 MHz presented by \citet{sirothia13}. (B) LOFAR full bandwidth HBA image at 138 MHz. (C) Reprocessing of the GMRT data used by \citet{sirothia13} at 325 MHz. (D) GMRT image at 153 MHz. All images are scaled to provide the same field of view. The restoring beams are indicated in the bottom left corner of the images.}
\label{ddrgcompimage}
\end{figure*}

\subsection{Remnant emission in 3C452}
\label{noddrg}

Until recently, 3C452 was thought to be a prototypical example of an FR II type radio galaxy with its strong hotspots and a classical morphology. This assertion was supported by observations presented by \citet{nandi10} who searched for remnant lobes using the Giant Metrewave Radio Telescope (GMRT; \citealp{swarup91}), but found no evidence for episodes of previous AGN activity. The discovery by \citet{sirothia13} that it may instead be a DDRG was therefore unexpected, with potentially significant consequences for models of radio galaxies, the unification of radio-loud AGN, and their application in cosmological studies. The steep spectrum of remnant lobes makes low-frequency observations such as those used within this paper ideal for the study of such emission. Assuming a (very) conservative estimate for the spectral index for the outer lobes of $\alpha = 0.5$, correcting for the smaller LOFAR beam size, and extrapolating the 3 sigma contour of \citet{sirothia13} back to 138 MHz, the faintest detected remnant emission should have a surface brightness of around 6 mJy beam$^{-1}$. If we assume the spectral index of 2.3 given by \citeauthor{sirothia13}, this increases to 27 mJy beam$^{-1}$, with brightest remnant regions being around 48 mJy beam$^{-1}$. This is well above the $<1$ mJy RMS noise of the LOFAR images and so should be clearly visible but, as was noted in Section \ref{morphology}, this emission is not detected in our observations (Figure \ref{ddrgcompimage}).

In order to determine the cause of this missing emission, we reprocessed the 325 MHz GMRT data originally used by \citet{sirothia13} along with an additional observation at 153 MHz available in the GMRT archive under project code 13SPA01 (Table \ref{gmrtobsdetails}). We reduced these archival observations using the \textsc{spam} data reduction package \citep{intema09}. After initial flagging of strong radio frequency interference (RFI) and dead antennas, flux, bandpass and instrumental phase calibrations were derived from the single scan on 3C48 (adopting the flux standard as defined in \citealp{scaife12}) and subsequently applied to the target field visibilities. Several iterations of self-calibration and wide-field imaging were started off with an initial phase-only gain calibration of the target field against a simple local sky model derived from NVSS, WENSS and VLSS. Each iteration also included flagging of weaker but more abundant RFI. The resulting images were corrected for primary beam attenuation, small astrometric offsets, and gain errors due to system temperature differences between flux calibrator and target field. From the resulting images shown in panels C and D of Figure \ref{ddrgcompimage} we again see that the putative outer lobes are not detected at either frequency.

Additional information can also be gained by considering the integrated flux of the source. Figure \ref{3C452integratedflux} shows the integrated spectrum of 3C452 between 74 MHz and 1.4 GHz with the GMRT images presented within this paper and by \citet{sirothia13}, alongside LOFAR and archival data. We see that within the errors, all measurements fall along a power law with the exception of the inner lobes of \citet{sirothia13}. This is further reinforced by the fact that this value can be bought back into agreement when the flux from both the inner and outer lobes are combined, suggesting that flux from the inner lobes has at some point during the reduction process been shifted to form the outer lobes.

Given that the remnant emission should be easily observable at low frequencies but is not seen in either the LOFAR, GMRT 150 MHz or archival observations (e.g. \citealp{kassim07}, at 74 MHz), that we are unable to reproduce the additional diffuse structure presented by \citet{sirothia13} in the GMRT 330 MHz map, and that the integrated flux can be brought back into agreement when the integrated flux of the inner and outer lobes are combined, we conclude that the most probable cause of the observed outer lobes is due to artefacts introduced during the self-calibration and/or the imaging of the source. We therefore conclude that 3C452 is not a DDRG, but is a standard FR II as was originally assumed.

\begin{table}
\centering
\caption{Summary of LOFAR maps by frequency}
\label{lofarmapdetails}
\begin{tabular}{lcccc}
\hline
\hline
Source&Frequency&Off-source RMS&\multicolumn{2}{c}{Integrated flux (Jy)}\\
&(MHz)&(mJy beam$^{-1}$)&Lobe 1&Lobe 2\\
\hline
3C452&116.9&1.46&44.27&41.65\\
&121.8&1.41&43.39&40.96\\
&124.7&1.37&42.56&40.67\\
&128.6&1.28&41.87&39.90\\
&132.5&1.21&41.28&39.55\\
&136.4&1.23&40.06&38.88\\
&140.3&1.05&39.71&38.49\\
&152.0&1.08&38.07&37.10\\
&155.9&1.00&37.64&36.66\\
&159.9&0.947&37.15&36.30\\

3C223&51.6&12.4&19.00&15.41\\
&118&1.08&13.69&11.0\\
&125&0.947&12.89&10.28\\
&133&0.910&12.03&9.76\\
&147&0.784&10.88&8.85\\
&154&0.840&10.38&8.40\\
&161&0.834&9.63&7.84\\
\hline

\end{tabular}

\vskip 5pt
\begin{minipage}{8.5cm}
`Frequency' refers to the frequency of the map, `Off-source RMS' refers to RMS noise measured over a large region well away from the source. `Integrated flux' values are listed at each frequency for the two lobes where `Lobe 1' refers to the northern and eastern and `Lobe 2' to the southern and western lobes of 3C223 and 3C452 respectively.
\end{minipage}
\end{table}

\subsection{Departure from equipartition}
\label{equipartition}

Equipartition between a source's particle and magnetic field energy densities is a common assumption made in the study of radio galaxies in order to provide an estimate of the magnetic field strength but, as was described in Section \ref{energertics}, the addition of X-ray observations allows the value of the magnetic field strength to be determined free from this constraint. While for an intrinsically steep initial electron energy distribution (discussed further in Section \ref{energeticsdicussion}) the absolute magnetic field strength increases by $\sim$60 per cent compared to \citet{croston04, croston05}, its relative strength remains much lower compared to equipartition irrespective of model, being only $\sim$50 and $\sim$75 per cent of the equipartition values for 3C452 and 3C223 respectively.

Deviation from equipartition is observed in similar FR II sources by \citet{croston04, croston05} and in 3C452 by \citet{shelton11}, who find $B \approx 0.3B_{eq}$ using inverse-Compton measurements independent of those of \citet{croston05}. Such a difference therefore comes as no surprise, but as a consequence the age of sources as derived from their spectrum are also affected. Spectral ageing, the preferential cooling of high energy electrons due to synchrotron and inverse-Compton losses \citep{kardashev62, pacholczyk70, jaffe73, tribble93}, has long been used as a method of determining the age of radio galaxies and is a practice which continues to this day (e.g. \citealp{carilli91, jamrozy05, orru10, heesen14, brienza16}), but there is often a large discrepancy between the ages determined in this way and those determined from a dynamical view point \citep{eilek96a, harwood13, harwood15}. In many cases, a simple solution to this problem is a weaker magnetic field strength, leading to lower radiative losses and older spectral ages. For example, in the case of 3C438 a magnetic field strength of only $B \approx 0.5B_{eq}$ is required to bring the spectral age back into agreement with the dynamical age. However, in order to determine whether these departures from equipartition are able to provide the magnetic field strength required to bring the two ages into agreement for the population as a whole, a detailed ageing study of a large sample of radio galaxies where both X-ray and radio observations are available is required.

It is of course not feasible to use X-ray observations for all sources studied but, while the absolute magnetic field strength for both 3C452 and 3C223 is plausibly greater than previously estimated, the increased total energy density of the lobes also changes the equipartition value as a natural consequence of the assumed relationship between the two values. Consequently, the ratio between the equipartition value and that found from synchrotron/inverse-Compton model fitting for both 3C452 and 3C223 remains roughly constant, varying on the order of only 10 per cent. The results of \citet{croston05}, who find a strong peak around a value of $0.7B_{eq}$, are therefore likely to be robust and provide a better estimate of the magnetic field strength in FR IIs than equipartition alone. Future studies using low-frequency radio surveys and the increased availability of archival X-ray observations will further refine this value, but these weaker field strengths may go at least part way to resolving the outstanding problem of explaining the spectrum of powerful radio galaxies. We suggest that such deviations should be carefully considered in future studies of FR IIs in order to provide a more accurate estimate of a source's age.

 \begin{table}
\centering
\caption{Model fitting results}
\label{lofaricres}
\begin{tabular}{llcccc}
\hline
\hline
Source&Lobe&\multicolumn{2}{c}{Magnetic field (nT)}&Energy density\\
&&B$_{iC}$&B$_{eq}$&($10^{-12}$ J m$^{-3}$)\\
\hline
3C452&Total&0.45&0.87&1.2\\
3C223&Northern&0.36&0.45&0.28\\
&Southern&0.32&0.45&0.32\\
\hline
\end{tabular}
\vskip 5pt
\begin{minipage}{8.5cm}
Synchrotron/inverse-Compton model fitting results assuming an intrinsically steep injection index of $\alpha_{inj} =$ 0.85 and 0.71 for 3C452 and 3C223 respectively. `Magnetic field' and `Energy density' give the results for the region listed in the `Lobe' column, where B$_{iC}$ and B$_{eq}$ give the magnetic field strength as derived from inverse-Compton measurements and equipartition respectively.
\end{minipage}
\end{table}

\subsection{Energetics and lobe pressures}
\label{energeticsdicussion}

As the loss time scale, $\tau$, of an electron radiating via the synchrotron process scale as \begin{equation}\label{elosses}\tau = \frac{E}{dE/dt} \propto 1/E \propto 1/\nu^{2} \end{equation} it is commonly assumed that for all but the oldest sources (i.e. those on the order of a few 100 Myr or more), the low frequency spectrum of radio galaxy lobes maintain their original power law form, providing a proxy for the initial electron energy distribution. Previous studies of 3C452 and 3C223 (e.g. \citealp{nandi10} for 3C452; \citealp{orru10} for 3C223) suggest that both sources fall well within the required age range for this assumption to be valid; however, the flux measurements presented in Section \ref{energertics} show that the integrated spectrum at low frequencies is much steeper than the traditionally assumed injection index value of  $\alpha_{inj} \approx$ 0.5 ($\delta \approx$ 2.0)\footnote{It is interesting to note that these values also agree with the average low-frequency (178 - 750 MHz) integrated spectral index of the 3CRR sample \citep{laing83} and those of \citet{nandi10} who find an injection index of 0.78 for 3C452. There is also some evidence that the jets of FR Is are slightly (but significantly) steeper then the expected $\alpha_{inj} = 0.5$ \citep{young05, laing13}.}.

\begin{figure}
\centering
\includegraphics[angle=0,width=7.7cm]{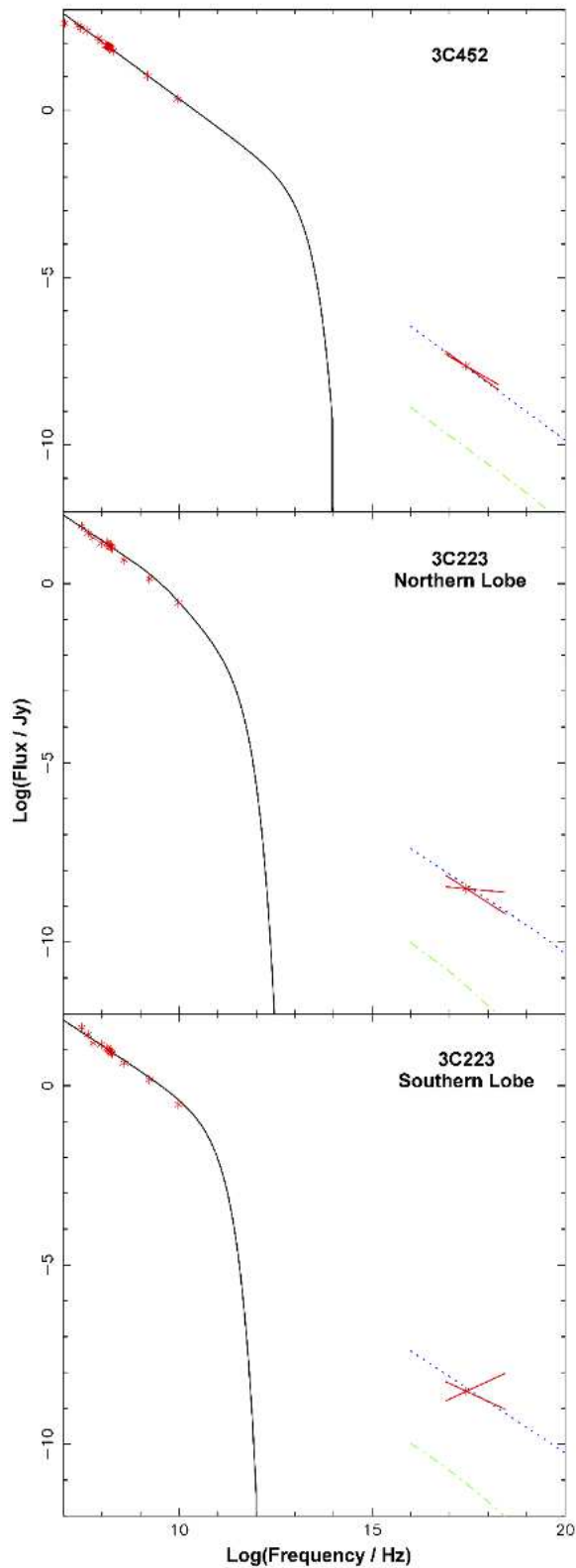}
\caption{Synchtotron/inverse-Compton model fitting for 3C452 (top), and the northern (middle) and southern (bottom) lobes of 3C223. Red stars indicate the radio measurements fitted with a synchrotron spectrum (solid black line). The red bow tie indicates constraints placed by X-ray measurements with 1 sigma errors on the fitted inverse-Compton model (blue dotted line). The green dash-dot line shows the expected location of the X-ray in the case of synchrotron self-Compton.}
\label{synchicfit}
\end{figure}

\begin{figure*}
\centering
\includegraphics[angle=-90,width=12.0cm]{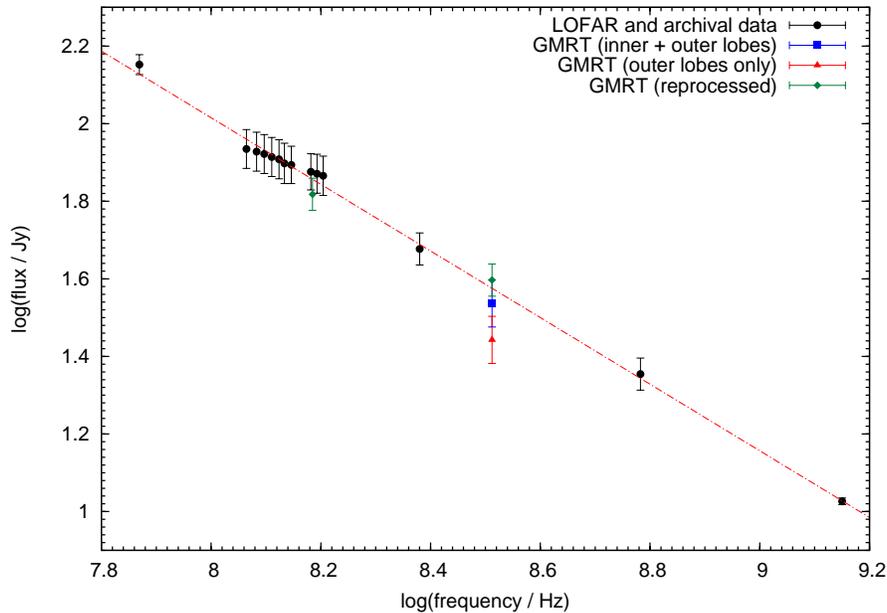}
\caption{Integrated flux of 3C452 between 74 MHz and 1.4 GHz. Black circles denote the LOFAR and archival data used in the synchrotron/inverse-Compton model fitting, with the dash-dot line representing the best fitting regression to this data. The red triangle indicates the flux of the inner lobes, and the blue square the combined flux of the inner and outer lobes as presented by \citet{sirothia13}. The green diamonds show the values of the reprocessed GMRT data presented in this paper. Note that the upper error bar of the combined inner and outer lobe GMRT data passes through the reprocessed GMRT data point.}
\label{3C452integratedflux}
\end{figure*}

One possible cause of this observed steeper than expected injection index is that the measured spectrum at LOFAR frequencies is not representative of the underlying electron energy distribution. This is expected if the source undergoes time dependent adiabatic and radiative losses throughout its life (e.g. \citealp{murgia99}) and is supported by simulations of FR II radio galaxies that suggest the integrated spectrum at these frequencies may be driven to steeper values due to changes in the magnetic field strength, mixing of electron populations, and the environment as a function of time \citep{kapinska15}. Under such conditions, the spectrum would not provide a good proxy for the initial electron distribution and typical models of spectral ageing which are applied to the integrated flux (e.g. CI models, \citealp{pacholczyk70}) would not provide an accurate description of the particle energy spectrum leading to an unreliable measure of a source's intrinsic age. Indeed, the fact that for 3C223 the X-ray spectral index (constrained by the bow ties of Figure \ref{synchicfit}) appears flatter than the radio spectral index indicates that the electron energy distribution flattens towards lower energies. The X-ray photons are produced by electrons/positrons with $\gamma \sim 1000$, a factor of $\gtrsim 2$ less than those responsible for the radio emission in the LOFAR HBA band, and comparable to those radiating in the LBA band, assuming a magnetic field strength $\sim 0.5$~nT.

An alternative explanation is that a break in the spectrum occurs at a much lower frequencies than is classically expected for an FR II radio galaxy. If this is the case, the spectrum may eventually flatten to injection index values closer to the expected value of 0.5 and the total energy content of the lobes will be in better agreement with those derived by \citet{croston04, croston05}. The low frequency archival observations used in the model fitting suggest that the integrated spectrum of both 3C452 and 3C223 remains steep down to $\approx$20 MHz, although the uncertainty on these measurements does allow for some moderate curvature to be present. Observations at the very lowest frequencies ($\approx$10 MHz; \citealp{braude70, laing80}), are limited by their large errors bars ($\approx 25$ percent) and so it is not possible to conclusively determine whether a break at very low frequencies is present in the spectrum. MHD modelling suggests that producing integrated spectra with such a low-frequency break is possible (Hardcastle et al., in preparation), but requires a spectrum not typically described by models of spectral ageing and there is no easy physical interpretation as to why such a break would occur.

The most intuitive physical explanation for the steep low-frequency electron energy distribution observed in both 3C452 and 3C223 is a weaker than expected shock due to a slow moving jet, as may plausibly be the case for less powerful objects such as 3C438 \citep{harwood15}. Theres has been some suggestion of a correlation between jet power and injection index \citep{konar13} with weaker shocks resulting in a steeper initial electron energy distribution; however, the strong, bright, compact hotspots observed in both 3C452 and 3C223 mean that this is unlikely to be the case. It is therefore more plausible that the emission results from a fast jet. It is theoretically possible to produce a high injection index under certain conditions through models such as those presented by \citet{konar13}, although further investigation is required to determine if this is intrinsically the case.

Support in favour of an intrinsically steep injection index also comes from the spatially resolved spectra of FR II radio lobes. \citet{harwood13, harwood15}, who consider the spectrum of 4 other FR II radio galaxies on smaller spatial scales, find that in all cases the injection index is also steeper than previously assumed. These studies attempt to account for low-frequency curvature when determining the injection index (hence electron energy distribution) of a source and, due to being well resolved, should be relatively unaffected by the superposition of spectra due to limited resolution. Injection index values found in this way agree well with the low-frequency spectral indices presented in this paper and have proved robust against simulations (e.g. \citealp{stroe14}). We note that preliminary tests (which will form part of paper II of this series) also find a similar distribution using the methods of \citet{harwood13, harwood15} implemented in the \textsc{brats}\footnote{http://www.askanastronomer.co.uk/brats} software package; however, a similar assumption is made that the injection index is constant over the lifetime of the source. As simulations of FR IIs are currently unable to trace the spatial evolution and mixing of the underlying electron population, it is not yet clear on what scales such variations exist and to what extent this impacts upon the determination of injection index values determined in this way, but further developments should make this possible over the next few years.

A physically plausible argument can therefore be made for an intrinsically steep injection index which, if correct, would have a significant impact on our understanding of the energetics of FR II galaxies. Assuming no significant flattening of the spectrum at frequencies $\lesssim$20 MHz, we can infer that a greater electron population is present at very low energies than previously assumed, resulting in the total energy content of the lobes increasing by a factor of between 2 and 5 compared to previous findings (Table \ref{lofaricres}). Using the standard equation for a relativistic plasma, $P_{lobe}=U/3$, where $P_{lobe}$ is the lobe pressure and $U$ is the total energy density, we see a significant change in the estimated internal pressure of the lobes (Table \ref{lobepressures}). Comparing these values to the external pressures at the tip of the lobes derived by \citet{shelton11} of  $1.1 \pm 0.1 \times 10^{-13}$ Pa for 3C452 and by \citet{croston04} of $9.6 \pm^{38.3}_{8.8} \times 10^{-14}$ Pa for 3C223, we see that they are overpressured by a factor of about 4 for 3C452, with the lobes of 3C223 being approximately in pressure balance with the external medium.

\begin{table}
\centering
\caption{Overview of GMRT observations of 3C452}
\label{gmrtobsdetails}
\begin{tabular}{ccccc}
\hline
\hline
Frequency&Target TOS&Calibrator&RMS (mJy&Resolution\\
(MHz)& (mins)&&beam$^{-1})$&(arcsec)\\
\hline
153&370&3C48&4.5&22.5 16.5\\
325&244&3C48&1.7&11.3 10.0\\
\hline
\end{tabular}
\vskip 5pt
\begin{minipage}{8.4cm}
Details of the archival GMRT observations for 3C452. `Frequency' lists the central frequency at which the final images were made. `Target TOS' list the observation length for the target and `Calibrator' the source observed for flux calibration purposes. `RMS' lists the off-source noise and `Resolution' the restoring beam of the final images.
\end{minipage}
\end{table}

For these revised values, our results for 3C452 broadly agree with those of \citet{shelton11} who also find that the lobes are overpressured; however, our revised estimate of the total energy content increases the internal overpressure by a factor of 2 compared to their findings. Using the standard equation relating the Mach number of the shock, $M$, to the internal to external pressure ratio \citep{longair11} and rearranging in terms $M$ we find \begin{equation}\label{machnumber}M = \frac{\sqrt{2 ((\gamma_{sh}+1) P_{ext} + (\gamma_{sh}-1) P_{int} )}} {2 \sqrt{\gamma_{sh}} \sqrt{P_{int}}}\end{equation} where $P_{ext}$ and $P_{int}$ are the external and internal pressures and, for a monatomic gas, $\gamma_{sh} = 5/3$ is the ratio of specific heats. This gives a Mach number of $M = 1.8$ implying the lobes are expending supersonically and driving a shock as the lobes push through the external medium. Although such a relatively weak shock is unlikely to be observable with current X-ray instruments, a temperature increase is observed in the external medium just beyond the tip of the lobes (region 3 of \citealp{shelton11}). We can therefore be fairly confident that the lobes are both intrinsically overpressured, and expanding through the external medium supersonically.

For 3C223, our results once again agree within errors to those of \citet{croston04} with the tip of the lobe being at minimum in pressure balance with the external medium. While the results presented in this paper suggest that the internal pressure of the lobes may be significantly higher than those derived by \citet{croston04}, the large uncertainty on the X-ray measurements from which the external pressure is derived means that both the overpressured and pressure balance cases are plausible. If one is to obtain the ratio of internal to external pressures to a higher degree of accuracy (e.g. to test the case of overpressured lobes) then improved X-ray measurements are required. While we cannot therefore rule out either case, this pressure increase supports the interpretation of \citet{croston04} that the source does, at minimum, have the required internal pressure to support the observed lobes. This is in line with other recent investigations in to the pressure balance of FR IIs (Ineson et al., in prep) which suggest that, while some exceptions are observed (e.g. 3C444, \citealp{croston11}), the FR II population as a whole is at minimum in pressure balance with the external medium when the magnetic field strength is derived from synchrotron/inverse-Compton model fitting.

From the results presented in this paper, it is clear that a revision in our understanding of radio galaxy lobes at LOFAR frequencies is required. Either the low-energy electron population is much greater than previously assumed, or new models describing the emission from the lobes of radio galaxies are required if we are to accurately determine the dynamics and energetics of FR-II radio galaxies.

\begin{table}
\centering
\caption{Summary of lobe pressures}
\label{lobepressures}
\begin{tabular}{llccc}
\hline
\hline
Source&Lobe&$P_{lobe}$ (Pa)&$P_{ext}$ (Pa)&$Ratio$\\
\hline
3C452&Total&$4.0 \times 10^{-13}$&$1.11 \times 10^{-13}$&3.60\\
3C223&Northern&$9.3 \times 10^{-14}$&$9.6 \times 10^{-14}$&0.97\\
&Southern&$1.1 \times 10^{-13}$&$9.6 \times 10^{-14}$&1.15\\
\hline
\end{tabular}
\vskip 5pt
\begin{minipage}{8.5cm}
Table of derived lobe pressures for 3C452 and 3C223, assuming the steep injection indices discussed in Section \ref{energeticsdicussion}. `$P_{lobe}$' refers to the pressure of the lobe listed in the `Lobe' column. `$P_{ext}$' is the external pressure taken from \citet{shelton11} for 3C452 and \citet{croston04} for 3C223. `Ratio' is the ratio of the lobe to external pressures.
\end{minipage}
\end{table}

\section{Conclusions}
\label{conclusions}

In this paper we have presented a low-frequency study of two FR II sources, 3C452 and 3C223. Using LOFAR and complementary archival observations we have explored the morphology of these sources and, by constraining the integrated spectrum between $\sim$10 and 8000 MHz, have investigated their energetics. We have used synchrotron/inverse-Compton fitting to test 2 possible models of the low energy electron population and, for the case of an intrinsically steep injection index, provided revised estimates of the magnetic field strength, total energy content and pressure of the radio lobes. We went on to discuss the impact of these findings on the pressure balance with the external medium. The key points made within this paper are as follows:

\begin{enumerate}
\item The morphology of 3C452 is that of a standard FR II type radio galaxy, contrary to the DDRG classification of \citet{sirothia13}.\\
\item In order to fit the low-frequency spectrum, either the injection index must be steeper than previously assumed or a break must be present at very low frequencies.\\
\item For an intrinsically steep initial electron energy distribution, we provide revised total energy content estimates which are between a factor of 2 and 5 greater than previous findings.\\
\item For these revised values, we find that the magnetic field strength of both sources is greater by around 60 per cent compared previous estimates, with values of 0.45 nT for 3C452 and 0.36 and 0.32 nT for the northern and southern lobes of 3C223 respectively.\\
\item The ratio between equipartition fields and those derived through synchrotron/inverse-Compton model fitting remains consistent with previous findings at 0.50 and 0.75 for 3C452 and 3C223 respectively.\\
\item We suggest that the observed departure from equipartition may, in some cases, provide a solution to the spectral versus dynamical age disparity problem.\\
\end{enumerate}

We therefore conclude that the morphology of both 3C452 and 3C223 are that of classical FR II galaxies, but with an integrated low-frequency spectrum steeper than has previously been assumed. We will further investigate these findings and the spectrum of these sources on resolved spatial scales in the second paper in this series.

\section{Acknowledgements}
\label{acknowledgements}
We wish to thank the anonymous referee, Francesco de Gasperin, Clive Tadhunter, Peter Barthel and George Heald for their constructive comments which have helped improve this paper. JJH wishes to thank the Netherlands Institute for Radio Astronomy (ASTRON) for a postdoctoral fellowship. This research was partly funded by the European Research Council under the European Union's Seventh Framework Programme (FP/2007-2013)/ERC Advanced Grant RADIOLIFE-320745. JHC, MJH and PNB are grateful for support from the Science and Technology Facilities Council under grants ST/M001326/1, ST/M001008/1 and ST/M001229/1. HTI acknowledges support from the National Radio Astronomy Observatory, a facility of the National Science Foundation operated under cooperative agreement by Associated Universities, Inc. AS is grateful for support from the European Research Council Advanced Grant 267697 4 $\pi$ sky: Extreme Astrophysics with Revolutionary Radio Telescopes. The Low Frequency Array was designed and constructed by ASTRON (Netherlands Institute for Radio Astronomy), and has facilities in several countries, that are owned by various parties (each with their own funding sources), and that are collectively operated by the International LOFAR Telescope (ILT) foundation under a joint scientific policy. We thank the staff of the GMRT that made these observations possible. GMRT is run by the National Centre for Radio Astrophysics of the Tata Institute of Fundamental Research. This research has made use of the NASA/IPAC Extragalactic Database (NED), which is operated by the Jet Propulsion Laboratory, California Institute of Technology, under contract with the National Aeronautics and Space Administration.

\bibliographystyle{mn2e}

\def\newblock{\hskip .11em plus .33em minus .07em}
\bibliography{lofar_energetics}

\newcommand{\noop}[1]{}
\begin{thebibliography}{}

\bibitem[\protect\citeauthoryear{Alexander \& Leahy}{Alexander \&
  Leahy}{1987}]{alexander87}
Alexander P.,  Leahy J.~P.,  1987, MNRAS, 225, 1

\bibitem[\protect\citeauthoryear{Bower, Benson, Malbon, Helly, Frenk, Baugh,
  Cole \& Lacey}{Bower et~al.}{2006}]{bower06}
Bower R.~G.,  Benson A.~J.,  Malbon J.~C.,  Helly J.~C.,  Frenk C.~S.,  Baugh
  C.~M.,  Cole S.,    Lacey C.~G.,  2006, MNRAS, 370, 645

\bibitem[\protect\citeauthoryear{{Braude}, {Lebedeva}, {Megn}, {Ryabov} \&
  {Zhouck}}{{Braude} et~al.}{1970}]{braude70}
{Braude} S.~Y.,  {Lebedeva} O.~M.,  {Megn} A.~V.,  {Ryabov} B.~P.,    {Zhouck}
  I.~N.,  1970, Astrophys. Lett., 5, 129

\bibitem[\protect\citeauthoryear{Brienza et~al.,}{Brienza
  et~al.}{2016}]{brienza16}
Brienza M.  et~al., 2016, A\&A, 585, 29

\bibitem[\protect\citeauthoryear{{Burbidge}}{{Burbidge}}{1956}]{burbidge56}
{Burbidge} G.~R.,  1956, ApJ, 124, 416

\bibitem[\protect\citeauthoryear{Carilli, Perley, Dreher \& Leahy}{Carilli
  et~al.}{1991}]{carilli91}
Carilli C.,  Perley R.,  Dreher J.,    Leahy J.,  1991, ApJ, 383, 554

\bibitem[\protect\citeauthoryear{Cohen, Lane, Cotton, Kassim, Lazio, Perley,
  Condon \& Erickson}{Cohen et~al.}{2007}]{cohen07}
Cohen A.~S.,  Lane W.~M.,  Cotton W.~D.,  Kassim N.~E.,  Lazio T. J.~W.,
  Perley R.~A.,  Condon J.~J.,    Erickson W.~C.,  2007, AJ, 134, 1245

\bibitem[\protect\citeauthoryear{Condon, Broderick, Seielstad, Douglas \&
  Gregory}{Condon et~al.}{1994}]{condon94}
Condon J.~J.,  Broderick J.~J.,  Seielstad G.~A.,  Douglas K.,    Gregory
  P.~C.,  1994, AJ, 107, 1829

\bibitem[\protect\citeauthoryear{Condon, Cotton, Greisen, Yin, Perley, Taylor
  \& Broderick}{Condon et~al.}{1998}]{condon98}
Condon J.~J.,  Cotton W.~D.,  Greisen E.~W.,  Yin Q.~F.,  Perley R.~A.,  Taylor
  G.~B.,    Broderick J.~J.,  1998, AJ, 115, 1693

\bibitem[\protect\citeauthoryear{Croston, Birkinshaw, Hardcastle \&
  Worrall}{Croston et~al.}{2004}]{croston04}
Croston J.~H.,  Birkinshaw M.,  Hardcastle M.~J.,    Worrall D.~M.,  2004,
  MNRAS, 353, 879

\bibitem[\protect\citeauthoryear{Croston, Hardcastle, Harris, Belsole,
  Birkinshaw \& Worrall}{Croston et~al.}{2005}]{croston05}
Croston J.~H.,  Hardcastle M.~J.,  Harris D.~R.,  Belsole E.,  Birkinshaw M.,
   Worrall D.~M.,  2005, ApJ, 626, 733

\bibitem[\protect\citeauthoryear{Croston, Hardcastle, Mingo, Evans, Dicken,
  Morganti \& Tadhunter}{Croston et~al.}{2011}]{croston11}
Croston J.~H.,  Hardcastle M.~J.,  Mingo B.,  Evans D.~A.,  Dicken D.,
  Morganti R.,    Tadhunter C.~N.,  2011, ApJ, 734, L28

\bibitem[\protect\citeauthoryear{Croton et~al.,}{Croton
  et~al.}{2006}]{croton06}
Croton D.~J.  et~al., 2006, MNRAS, 365, 11

\bibitem[\protect\citeauthoryear{Eilek}{Eilek}{1996}]{eilek96a}
Eilek J.~A.,  1996, in Hardee P.~E.,  Bridle A.~H.,   Zensus J.~A.,  eds,  ASP
  Conference Series Vol. 100, \emph{`Energy transport in radio galaxies and
  quasars'}.

\bibitem[\protect\citeauthoryear{{Fabian}}{{Fabian}}{2012}]{fabian12}
{Fabian} A.~C.,  2012, ARAA, 50, 455

\bibitem[\protect\citeauthoryear{Fanaroff \& Riley}{Fanaroff \&
  Riley}{1974}]{fanaroff74}
Fanaroff B.~L.,  Riley J.~M.,  1974, MNRAS, 167, 31P

\bibitem[\protect\citeauthoryear{Godfrey \& Shabala}{Godfrey \&
  Shabala}{2015}]{godfrey15}
Godfrey L.,  Shabala S.,  2015, Mutual distance dependence drives the observed
  jet power - radio luminosity scaling relations in radio galaxies, submitted

\bibitem[\protect\citeauthoryear{{Godfrey} et~al.,}{{Godfrey}
  et~al.}{2009}]{godfrey09}
{Godfrey} L.~E.~H.  et~al., 2009, ApJ, 695, 707

\bibitem[\protect\citeauthoryear{Hardcastle, Alexander, Pooley \&
  Riley}{Hardcastle et~al.}{1998}]{hardcastle98a}
Hardcastle M.~J.,  Alexander P.,  Pooley G.~G.,    Riley J.~M.,  1998, MNRAS,
  296, 445

\bibitem[\protect\citeauthoryear{Hardcastle, Birkinshaw \& Worrall}{Hardcastle
  et~al.}{1998}]{hardcastle98}
Hardcastle M.~J.,  Birkinshaw M.,    Worrall D.~M.,  1998, MNRAS, 294, 615

\bibitem[\protect\citeauthoryear{Hardcastle \& Croston}{Hardcastle \&
  Croston}{2005}]{hardcastle05}
Hardcastle M.~J.,  Croston J.~H.,  2005, MNRAS, 363, 649

\bibitem[\protect\citeauthoryear{Hardcastle, Harris, Worrall, Birkinshaw, Laing
  \& H.}{Hardcastle et~al.}{2004}]{hardcastle04b}
Hardcastle M.~J.,  Harris D.~E.,  Worrall D.~M.,  Birkinshaw M.,  Laing R.~A.,
    H. B.~A.,  2004, ApJ, 612, 729

\bibitem[\protect\citeauthoryear{Hardcastle \& Worrall}{Hardcastle \&
  Worrall}{2000}]{hardcastle00}
Hardcastle M.~J.,  Worrall D.~M.,  2000, MNRAS, 319, 562

\bibitem[\protect\citeauthoryear{Harwood, Hardcastle \& Croston}{Harwood
  et~al.}{2015}]{harwood15}
Harwood J.~J.,  Hardcastle M.~J.,    Croston J.~H.,  2015, MNRAS, 454, 3403

\bibitem[\protect\citeauthoryear{Harwood, Hardcastle, Croston \&
  Goodger}{Harwood et~al.}{2013}]{harwood13}
Harwood J.~J.,  Hardcastle M.~J.,  Croston J.~H.,    Goodger J.~L.,  2013,
  MNRAS, 435, 3353

\bibitem[\protect\citeauthoryear{Heald et~al.,}{Heald et~al.}{2010}]{heald10}
Heald G.  et~al., 2010, in {ASTRON Netherlands Institute for Radio Astronomy}
  ed., \emph{`Proceedings of the ISKAF2010 Science Meeting'}. ISKAF2010.
p.~57

\bibitem[\protect\citeauthoryear{Heckman \& Best}{Heckman \&
  Best}{2014}]{heckman14}
Heckman T.~M.,  Best P.~N.,  2014, ARA\&A, 52, 589

\bibitem[\protect\citeauthoryear{Heesen, Croston, Harwood, Hardcastle \&
  Ananda}{Heesen et~al.}{2014}]{heesen14}
Heesen V.,  Croston J.~H.,  Harwood J.~J.,  Hardcastle M.~J.,    Ananda H.,
  2014, MNRAS, 439, 1364

\bibitem[\protect\citeauthoryear{Homan \& Wardle}{Homan \&
  Wardle}{1999}]{homan99}
Homan D.~C.,  Wardle J. F.~C.,  1999, AJ, 118, 1942

\bibitem[\protect\citeauthoryear{Intema, van~der Tol, Cotton, Cohen, van Bemmel
  \& R\"ottgering}{Intema et~al.}{2009}]{intema09}
Intema H.~T.,  van~der Tol S.,  Cotton W.~D.,  Cohen A.~S.,  van Bemmel I.~M.,
    R\"ottgering H. J.~A.,  2009, A\&A, 501, 1185

\bibitem[\protect\citeauthoryear{Isobe, Tashiro, Makishima, Iyomoto, Suzuki,
  Murakami, Mori \& Abe}{Isobe et~al.}{2002}]{isobe02}
Isobe N.,  Tashiro M.,  Makishima K.,  Iyomoto N.,  Suzuki M.,  Murakami M.~M.,
   Mori M.,    Abe K.,  2002, ApJ, 580, L111

\bibitem[\protect\citeauthoryear{Jaffe \& Perola}{Jaffe \&
  Perola}{1973}]{jaffe73}
Jaffe W.,  Perola G.,  1973, A\&A, 26, 423

\bibitem[\protect\citeauthoryear{Jamrozy, Machalski, Mack \& Klein}{Jamrozy
  et~al.}{2005}]{jamrozy05}
Jamrozy M.,  Machalski J.,  Mack K.-H.,    Klein U.,  2005, A\&A, 433, 467

\bibitem[\protect\citeauthoryear{Kapinska, Hardcastle, Jackson, An T.;~Baan \&
  Jarvis}{Kapinska et~al.}{2015}]{kapinska15}
Kapinska A.~D.,  Hardcastle M.,  Jackson C.,  An T.;~Baan W.,    Jarvis M.,
  2015, in Bourke. T.~L.  et~al., eds, \emph{`Proceedings of Advancing
  Astrophysics with the Square Kilometre Array'}. AASKA14.
p.~173

\bibitem[\protect\citeauthoryear{Kardashev}{Kardashev}{1962}]{kardashev62}
Kardashev N.~S.,  1962, AJ, 6, 317

\bibitem[\protect\citeauthoryear{Kassim et~al.,}{Kassim
  et~al.}{2007}]{kassim07}
Kassim N.~E.  et~al., 2007, ApJ, 172, 686

\bibitem[\protect\citeauthoryear{Kataoka \& Stawarz}{Kataoka \&
  Stawarz}{2005}]{kataoka05}
Kataoka J.,  Stawarz L.,  2005, ApJ, 622, 797

\bibitem[\protect\citeauthoryear{Konar \& Hardcastle}{Konar \&
  Hardcastle}{2013}]{konar13}
Konar C.,  Hardcastle M.~J.,  2013, MNRAS, 436, 1595

\bibitem[\protect\citeauthoryear{Konar, Saikai, Jamrozy \& Machalski}{Konar
  et~al.}{2006}]{konar06}
Konar C.,  Saikai D.~J.,  Jamrozy M.,    Machalski J.,  2006, MNRAS, 372, 693

\bibitem[\protect\citeauthoryear{Laing \& Bridle}{Laing \&
  Bridle}{2013}]{laing13}
Laing R.~A.,  Bridle A.~H.,  2013, MNRAS, 432, 1114

\bibitem[\protect\citeauthoryear{Laing \& Peacock}{Laing \&
  Peacock}{1980}]{laing80}
Laing R.~A.,  Peacock J.~A.,  1980, MNRAS, 190, 903

\bibitem[\protect\citeauthoryear{Laing, Riley \& Longair}{Laing
  et~al.}{1983}]{laing83}
Laing R.~A.,  Riley J.~M.,    Longair M.~S.,  1983, MNRAS, 204, 151

\bibitem[\protect\citeauthoryear{Leahy \& Perley}{Leahy \&
  Perley}{1991}]{leahy91a}
Leahy J.~P.,  Perley R.~A.,  1991, AJ, 102, 537

\bibitem[\protect\citeauthoryear{Longair}{Longair}{2011}]{longair11}
Longair M.~S.,  2011, High Energy Astrophysics.
Cambridge University Press

\bibitem[\protect\citeauthoryear{Machalski, Jamrozy \& Saikia}{Machalski
  et~al.}{2009}]{machalski09}
Machalski J.,  Jamrozy M.,    Saikia D.~J.,  2009, MNRAS, 395, 812

\bibitem[\protect\citeauthoryear{{Machalski}, {Kozie{\l}-Wierzbowska},
  {Jamrozy} \& {Saikia}}{{Machalski} et~al.}{2008}]{machalski08}
{Machalski} J.,  {Kozie{\l}-Wierzbowska} D.,  {Jamrozy} M.,    {Saikia} D.~J.,
  2008, ApJ, 679, 149

\bibitem[\protect\citeauthoryear{McNamara \& Nulsen}{McNamara \&
  Nulsen}{2012}]{mcnamara12}
McNamara B.~R.,  Nulsen P. E.~J.,  2012, New J. Phys., 14, 055023

\bibitem[\protect\citeauthoryear{Morganti, Fogasy, Paragi, Oosterloo \&
  Orienti}{Morganti et~al.}{2013}]{morganti13}
Morganti R.,  Fogasy J.,  Paragi Z.,  Oosterloo T.,    Orienti M.,  2013,
  Science, 341, 1082

\bibitem[\protect\citeauthoryear{Morganti, Oosterloo \& Tsvetanov}{Morganti
  et~al.}{1988}]{morganti88}
Morganti R.,  Oosterloo T.,    Tsvetanov Z.,  1988, AJ, 115, 915

\bibitem[\protect\citeauthoryear{Mullin, Hardcastle \& Riley}{Mullin
  et~al.}{2006}]{mullin06}
Mullin L.~M.,  Hardcastle M.~J.,    Riley J.~M.,  2006, MNRAS, 372, 113

\bibitem[\protect\citeauthoryear{{Murgia}, {Fanti}, {Fanti}, {Gregorini},
  {Klein}, {Mack} \& {Vigotti}}{{Murgia} et~al.}{1999}]{murgia99}
{Murgia} M.,  {Fanti} C.,  {Fanti} R.,  {Gregorini} L.,  {Klein} U.,  {Mack}
  K.-H.,    {Vigotti} M.,  1999, A\&A, 345, 769

\bibitem[\protect\citeauthoryear{Nandi, Pirya, Pal, Konar, Saikia \&
  Singh}{Nandi et~al.}{2010}]{nandi10}
Nandi S.,  Pirya A.,  Pal S.,  Konar C.,  Saikia D.~J.,    Singh M.,  2010,
  MNRAS, 404, 433

\bibitem[\protect\citeauthoryear{Offringa, de Bruyn, Biehl, Zaroubi, Bernardi
  \& Pandey}{Offringa et~al.}{2010}]{offringa10}
Offringa A.~R.,  de Bruyn A.~G.,  Biehl M.,  Zaroubi S.,  Bernardi G.,
  Pandey V.~N.,  2010, MNRAS, 405, 155

\bibitem[\protect\citeauthoryear{Orr\'u, Murgia, Feretti, Govoni, Giovannini,
  Lane, Kassim \& Paladino}{Orr\'u et~al.}{2010}]{orru10}
Orr\'u E.,  Murgia M.,  Feretti L.,  Govoni F.,  Giovannini G.,  Lane W.,
  Kassim N.,    Paladino R.,  2010, A\&A, 515, A50

\bibitem[\protect\citeauthoryear{Owen \& Ledlow}{Owen \& Ledlow}{1994}]{owen94}
Owen F.~N.,  Ledlow M.~J.,  1994, in Bicknell G.~V.,  Dopita M.~A.,   Quinn
  P.~J.,  eds,  ASP Conference Series Vol. 54, \emph{`The first Stromlo
  symposium: The physics of active galaxies'}.

\bibitem[\protect\citeauthoryear{Pacholczyk}{Pacholczyk}{1970}]{pacholczyk70}
Pacholczyk A.~G.,  1970, Radio astrophysics. Nonthermal processes in galactic
  and extragalactic sources.
San Francisco, Freeman

\bibitem[\protect\citeauthoryear{Pandey, van Zwieten, de Bruyn \&
  Nijboer}{Pandey et~al.}{2009}]{pandey09}
Pandey V.~N.,  van Zwieten J.~E.,  de Bruyn A.~G.,    Nijboer R.,  2009, in
  Saikia D.~J.,  Green D.~A.,  Gupta Y.,   Venturi T.,  eds,  ASP Conference
  Series Vol. 407, \emph{`The Low-Frequency Radio Universe'}.

\bibitem[\protect\citeauthoryear{Rau \& Cornwell}{Rau \&
  Cornwell}{2011}]{rau11}
Rau U.,  Cornwell T.~J.,  2011, A\&A, 532, 71

\bibitem[\protect\citeauthoryear{Rengelink, Tang, de Bruyn, Miley, Bremer \&
  R\"oettgering H. J. A.;~Bremer}{Rengelink et~al.}{1997}]{rengelink97}
Rengelink R.~B.,  Tang Y.,  de Bruyn A.~G.,  Miley G.~K.,  Bremer M.~N.,
  R\"oettgering H. J. A.;~Bremer M. A.~R.,  1997, A\&AS, 124, 259

\bibitem[\protect\citeauthoryear{Scaife \& Heald}{Scaife \&
  Heald}{2012}]{scaife12}
Scaife A. M.~M.,  Heald G.~H.,  2012, MNRAS, 423, L30

\bibitem[\protect\citeauthoryear{{Scheers}}{{Scheers}}{2011}]{scheers11}
{Scheers} L.~H.~A.,  2011, PhD thesis, University of Amsterdam

\bibitem[\protect\citeauthoryear{Schoenmakers, de Bruyn, R\"ottgering \&
  van~der Laan}{Schoenmakers et~al.}{2000}]{schoenmakers00}
Schoenmakers A.~P.,  de Bruyn A.~G.,  R\"ottgering H. J.~A.,    van~der Laan
  H.,  2000, MNRAS, 315, 371

\bibitem[\protect\citeauthoryear{Shelton, Hardcastle \& Croston}{Shelton
  et~al.}{2011}]{shelton11}
Shelton D.~L.,  Hardcastle M.~J.,    Croston J.~H.,  2011, MNRAS, 418, 811

\bibitem[\protect\citeauthoryear{Shulevski et~al.,}{Shulevski
  et~al.}{2015}]{shulevski15}
Shulevski A.  et~al., 2015, A\&A, 579, 27

\bibitem[\protect\citeauthoryear{Sikora, Stawarz \& Lasota}{Sikora
  et~al.}{2007}]{sikora07}
Sikora M.,  Stawarz L.,    Lasota J.-P.,  2007, ApJ, 658, 815

\bibitem[\protect\citeauthoryear{Sirothia, Gopal-Krishna \& Wiita}{Sirothia
  et~al.}{2013}]{sirothia13}
Sirothia S.~K.,  Gopal-Krishna S.,    Wiita P.~J.,  2013, ApJ, 765, L11

\bibitem[\protect\citeauthoryear{Spergel et~al.,}{Spergel
  et~al.}{2003}]{spergel03}
Spergel D.~N.  et~al., 2003, ApJS, 148, 175

\bibitem[\protect\citeauthoryear{Stroe, Harwood, Hardcastle \&
  R\"ottgering}{Stroe et~al.}{2014}]{stroe14}
Stroe A.,  Harwood J.~J.,  Hardcastle M.~J.,    R\"ottgering H. J.~A.,  2014,
  MNRAS, 445, 1213

\bibitem[\protect\citeauthoryear{Swarup, Ananthakrishnan, Kapahi, Rao,
  Subrahmanya \& Kulkarni}{Swarup et~al.}{1991}]{swarup91}
Swarup G.,  Ananthakrishnan S.,  Kapahi V.~K.,  Rao A.~P.,  Subrahmanya C.~R.,
    Kulkarni V.~K.,  1991, Curr. Sci., 60, 95

\bibitem[\protect\citeauthoryear{Tashiro et~al.,}{Tashiro
  et~al.}{1998}]{tashiro98}
Tashiro M.  et~al., 1998, ApJ, 499, 713

\bibitem[\protect\citeauthoryear{Tasse, van~der Tol, van Zwieten, van Diepen \&
  Bhatnagar}{Tasse et~al.}{2013}]{tasse13}
Tasse C.,  van~der Tol S.,  van Zwieten J.,  van Diepen G.,    Bhatnagar 2013,
  A\&A, 553, A105

\bibitem[\protect\citeauthoryear{Tribble}{Tribble}{1993}]{tribble93}
Tribble P.,  1993, MNRAS, 261, 57

\bibitem[\protect\citeauthoryear{{van Haarlem} et~al.,}{{van Haarlem}
  et~al.}{2013}]{haarlem13}
{van Haarlem} M.~P.  et~al., 2013, A\&A, 556, A2

\bibitem[\protect\citeauthoryear{van Weeren, Williams \& Tasse}{van Weeren
  et~al.}{2014}]{weeren14}
van Weeren R.~J.,  Williams W.~L.,    Tasse C.,  2014, ApJ, 793, 82

\bibitem[\protect\citeauthoryear{Wardle, Homan, Ojha \& Roberts}{Wardle
  et~al.}{1998}]{wardle98}
Wardle J. F.~C.,  Homan D.~C.,  Ojha R.,    Roberts D.~H.,  1998, Nature, 395,
  457

\bibitem[\protect\citeauthoryear{Worrall \& Birkinshaw}{Worrall \&
  Birkinshaw}{2000}]{worrall00}
Worrall D.~M.,  Birkinshaw M.,  2000, ApJ, 530, 719

\bibitem[\protect\citeauthoryear{{Young}, {Rudnick}, {Katz}, {DeLaney},
  {Kassim} \& {Makishima}}{{Young} et~al.}{2005}]{young05}
{Young} A.,  {Rudnick} L.,  {Katz} D.,  {DeLaney} T.,  {Kassim} N.~E.,
  {Makishima} K.,  2005, ApJ, 626, 748

\end{thebibliography}

\end{document}